\documentclass[journal]{IEEEtran}

\usepackage{booktabs} 
\usepackage{tabularx}
\usepackage[strict]{changepage}
\usepackage{calc}
\usepackage{mathtools,lipsum}
\usepackage{amsmath}
\usepackage{amssymb}
\usepackage{mathrsfs}
\usepackage{upgreek}
\usepackage{graphicx}
\usepackage{algorithmicx}
\usepackage{algorithm}
\usepackage{enumitem}
\usepackage[noend]{algpseudocode}
\usepackage{caption}
\usepackage{subcaption}
\usepackage{amsthm, times, amsfonts, cite}
\usepackage{tabularx}
\usepackage{color}

\newtheorem{theorem}{Theorem}
\newtheorem{lemma}{Lemma}
\newtheorem{corollary}{Corollary}

\usepackage[T1]{fontenc}
\usepackage[latin9]{inputenc}
\usepackage{array}
\usepackage{float}
\usepackage{amsmath}
\usepackage{amssymb}
\usepackage{graphicx}

\usepackage{amsmath}
\usepackage{algorithm}

\makeatletter
\def\BState{\State\hskip-\ALG@thistlm}
\makeatother

\begin{document}
\title{
 Video Streaming  in Distributed Erasure-coded  Storage Systems: Stall Duration Analysis}

\author{Abubakr O. Al-Abbasi and Vaneet Aggarwal \thanks{The authors are with the School of Industrial Engineering, Purdue University, West Lafayette IN 47907, email: \{aalabbas,vaneet\}@purdue.edu. This work was supported in
		part by the National Science Foundation under Grant no. CNS-1618335.
} }


\maketitle


\begin{abstract}
The demand for global video has been burgeoning across industries. With the expansion and improvement of  video-streaming services, cloud-based video is evolving into a necessary feature of any successful business for reaching internal and external audiences. This paper considers video streaming over distributed systems where the video segments are encoded using an erasure code for better reliability thus being the first work to our best knowledge that considers video streaming over erasure-coded distributed cloud systems. The download time of each coded chunk of each video segment is characterized and ordered statistics over the choice of the erasure-coded chunks is used to obtain the playback time of different video segments. Using the playback times, bounds on the moment generating function on the stall duration is used to bound the mean stall duration. Moment generating function based bounds on the ordered statistics are also used to bound the stall duration tail probability which determines the probability that the stall time is greater than a pre-defined number. These two  metrics, mean stall duration and the stall duration tail probability, are important quality of experience (QoE) measures for the end users.  Based on these metrics, we formulate an optimization problem to jointly minimize the convex combination of both the  QoE metrics averaged over all requests over the placement and access of the video content. The non-convex problem  is solved using an efficient iterative algorithm.  Numerical results show significant improvement in QoE metrics for cloud-based video as compared to the considered baselines.
\end{abstract}


%


\section{Introduction}

The demands of video streaming services have been skyrocketing over these years, with the global video streaming market expected to grow annually at a rate of  18.3\%  \cite{marketsandmarkets}.  With the proliferation and advancement of  video-streaming services, cloud-based video has become an imperative feature of any successful business. This can also be seen as IBM estimates cloud-based video will be a \$105 billion market opportunity by 2019 \cite{ibm}. In cloud storage systems, erasure coding has seen itself quickly emerged as a promising technique to reduce the storage cost for a given reliability as compared to the replicated systems \cite{2015_1,Dimakis:10}. It has been widely adopted in modern storage systems by companies like Facebook \cite{Sathiamoorthy13}, Microsoft \cite{Asure14}, and Google \cite{Fikes10}.  This paper considers video streaming when the content is placed on cloud servers, where erasure coding is used. The key quality of experience (QoE) metric for video streaming is the duration of  stalls at the clients. This paper gives bounds on the stall durations, and uses that to propose an optimized streaming service that minimizes average QoE for the clients.

In this paper, we consider two measures of QoE metrics in terms of stall duration. The first is the mean  stall duration.  Almost
every viewer can relate to the quality of experiences for watching videos being the stall duration and is thus one of the key focus in the studied streaming algorithms \cite{huang2015buffer,AnisTON}. The second is the probability that the stall duration is greater than a fixed number $x$, which determines the  stall duration tail probability.  It has been shown that in modern Web applications such as Bing, Facebook, and Amazon's retail platform, the long tail of latency is of particular concern, with $99.9$th percentile response times that are orders of magnitude worse than the mean \cite{T1,T2}. Thus, the QoE metric of stall duration tail probability becomes important. This paper characterizes an upper bound on both  QoE metrics.

We note that quantifying service latency for erasure-coded storage is an open problem \cite{MDS-Queue}, and so is tail latency \cite{Jingxian}. This paper takes a step forward and explores the notions for video streaming rather than video download. Thus, finding the exact QoE metrics is an open problem. This paper finds the bounds on the QoE metrics. The data chunk transfer time in practical systems follows a shifted exponential distribution \cite{Yu_TON,CS14} which motivates the choice that the service time distribution for each video server is a shifted exponential distribution. Further, the request arrival rates for each video is assumed to be Poisson. The video segments are encoded using an $(n,k)$ erasure code and the coded segments are placed on $n$ different servers. When a video is requested, the segments need to be requested from $k$ out of $n$ servers. Optimal strategy of choosing these $k$ servers would need a Markov approach similar to that in  \cite{MDS-Queue} and suffers from a similar state explosion  problem, because states
of the corresponding queuing model must encapsulate not
only a snapshot of the current system including chunk
placement and queued requests but also past history of how
chunk requests have been processed by individual nodes.

In this paper, we use the probabilistic scheduling proposed in \cite{Xiang:2014:Sigmetrics:2014,Yu_TON} to access the $k$ servers, where each possibility of $k$ servers is chosen with certain probability and the probability terms can be optimized. Using this scheduling mechanism, the random variables corresponding to the times for download of different video segments from each server are found. Using ordered statistics over the $k$ servers, the random variables corresponding to the playback time of each video segment are characterized. These are then used to find bounds on the mean stall duration and the stall duration tail probability. Moment generating functions of the ordered statistics of different random variables are used in the bounds. We note that the problem of finding latency for file download is very different from the video stall duration for streaming. This is because the stall duration accounts for download time of each video segment rather than only the download time of the last video segment. Further, the download time of segments are correlated since the download of chunks from a server are in sequence and the playback time of a video segment are dependent on the playback time of the last segment and the download time of the current segment. Taking these dependencies into account, this paper characterizes the bounds on the two QoE metrics. We note that for the special case when each video has a single segment, the bounds on mean stall duration and stall duration tail probability reduce to that for file download. Further, the bounds based on the approach in this paper have been shown to outperform the results for mean file download latency in \cite{Xiang:2014:Sigmetrics:2014,Yu_TON}.



The proposed framework provides a mathematical crystallization
of the engineering artifacts involved and illuminates
key system design issues through optimization of QoE. The average QoE metric over different  requests can be optimized over the placement of the video files,  the access of the video files from the servers, and the  bound parameters. The tradeoff in the two QoE metrics is captured by defining the objective function which is a convex combination of the two QoE metrics. Varying the parameter trading off the two metrics can be used to get a tradeoff region between the two metrics helping the system designer to choose an appropriate point. An efficient algorithm is proposed to solve the proposed non-convex problem. The proposed algorithm does an alternating optimization over the placement, access, and the bound parameters. The optimization over probabilistic scheduling access parameters help reduce the mean and tail of the stall durations by differentiating video files thus providing more flexibility as compared to choosing the lowest queue servers.

The sub-problems have been shown to have convex constraints and thus can be efficiently solved using iNner cOnVex
Approximation (NOVA)  algorithm proposed in \cite{scutNOVA}. The proposed algorithm is shown to converge to a local optimal.   Numerical results demonstrate significant improvement of QoE metrics as compared to the baselines.

 Today, cloud-based video does not use erasure coding. One of the key reason is  the additional decoding latency from multiple coded streams. Since the  computing has been growing exponentially \cite{Denning:2016:ELC:3028256.2976758}, it is only a matter of time when the computation of decoding will not limit the latencies in delay sensitive video streaming and the networking latency will govern the system designs. 
 Further, we note that replication is a special case of erasure coding. Thus, the proposed research using erasure-coded content on the servers can also be used when the content is replicated on the servers.

The key contributions of our paper include:
\begin{itemize}[leftmargin=0cm,itemindent=.5cm,labelwidth=\itemindent,labelsep=0cm,align=left]
\item This paper formulates video streaming over erasure-coded cloud storage system. 
\item The random variable corresponding to the download time of a chunk of each video segment from a server is characterized. Using ordered statistics, the random variable corresponding to the playback time of each video segment is found. These are further used to derive  upper bounds on the mean stall duration of the video and the video stall duration tail probability. 
\item The QoE metrics are used to formulate system optimization problems over the choice of the placement of video segments, probabilistic scheduling access policy and the bound parameters which are related to the moment generating function. Efficient iterative solutions are provided for these  optimization problems.
\item Numerical results show that the proposed algorithms
converges within a few iterations. Further, the QoE metrics are shown to have significant improvement as compared to the considered baselines. For instance, the mean stall duration for the proposed algorithm is 60\% smaller and the stall duration tail probability is orders of magnitude better as compared to random placement and projected equal access probability strategy.
\end{itemize}

The remainder of this paper is organized as follows. Section 2
provides related work for this paper. In Section 3, we describe
the system model used in the paper with a description of
video streaming over cloud storage. Section 4 derives expressions on the download and play times of the chunks which are used in  Sections 5 and 6 to find the upper bounds on the QoE metrics of the mean stall duration and video stall latency, respectively. Section 7 formulates the  QoE optimization problem as a weighted combination of the two QoE metrics  and  proposes the iterative algorithmic solution of this problem.   Numerical results are provided in Section 8. Section 9 concludes the paper.

\section{Related Work}

{\em Latency in Erasure-coded Storage: } To our best knowledge, however, while latency in erasure coded storage systems has been widely studied, quantifying exact latency for erasure-coded storage system in data-center network is an open problem. Prior works focusing on asymptotic queuing delay behaviors \cite{Bramson:10,Lu:10} are not applicable because redundancy factor in practical data centers typically remains small due to storage cost concerns. Due to the lack of analytic latency models, most of the literature is focused on reliable distributed storage system design, and latency is only presented as a performance metric when evaluating the proposed erasure coding scheme, e.g., \cite{AJX05,J06}, which demonstrate latency improvement due to erasure coding in different system implementations. Related design can also be found in data access scheduling \cite{SH07,TI10}, access collision avoidance \cite{ZA02}, and encoding/decoding time optimization \cite{WK} and there is also some work using the LT erasure codes to adjust the system to meet user requirements such as availability, integrity and confidentiality \cite{AG14}. 

Recently, there has been a number of attempts at finding latency bounds for an erasure-coded storage system \cite{Xiang:2014:Sigmetrics:2014,Yu_TON,CS14,Joshi:13,MDS-Queue}. The key scheduling approaches include {\em block-one-scheduling} policy that only allows the request at the head of the buffer to move forward \cite{MG1:12},  fork-join queue \cite{Makowski:89,Joshi:13} to request data from all server and wait for the first $k$ to finish, and the probabilistic scheduling \cite{Xiang:2014:Sigmetrics:2014,Yu_TON} that allows choice of every possible subset of $k$ nodes with certain probability. Mean latency and tail latency have been characterized in \cite{Xiang:2014:Sigmetrics:2014,Yu_TON} and \cite{Jingxian} respectively for a system with multiple files using probabilistic scheduling. This paper considers video streaming rather than file downloading. The metrics for video streaming does not only account for the end of the download of the video but also of the download of each of the segment. Thus, the analysis for the content download cannot be extended to the video streaming directly and the analysis approach in this paper is very different from the prior works in the area.

{\em Video Streaming over Cloud: } Servicing Video on Demand and Live TV Content from cloud servers have been studied widely \cite{lee2013vod,huang2011cloudstream,he2014cost,chang2016novel,oza2016implementation}. 
The placement of content and resource optimization over the cloud servers have been considered. To the best of our knowledge, reliability of content over the cloud servers have not been considered for video streaming applications. In the presence of erasure-coding, there are novel challenges to characterize and optimize the QoE metrics at the end user. Adaptive streaming algorithms have also been considered for video streaming \cite{chen2012amvsc,wang2013ames}, which are beyond the scope of this paper and are left for future work. 
\section{System Model}

\begin{figure}
\includegraphics[trim=0in 0in 0in .8in, clip, scale=0.35]{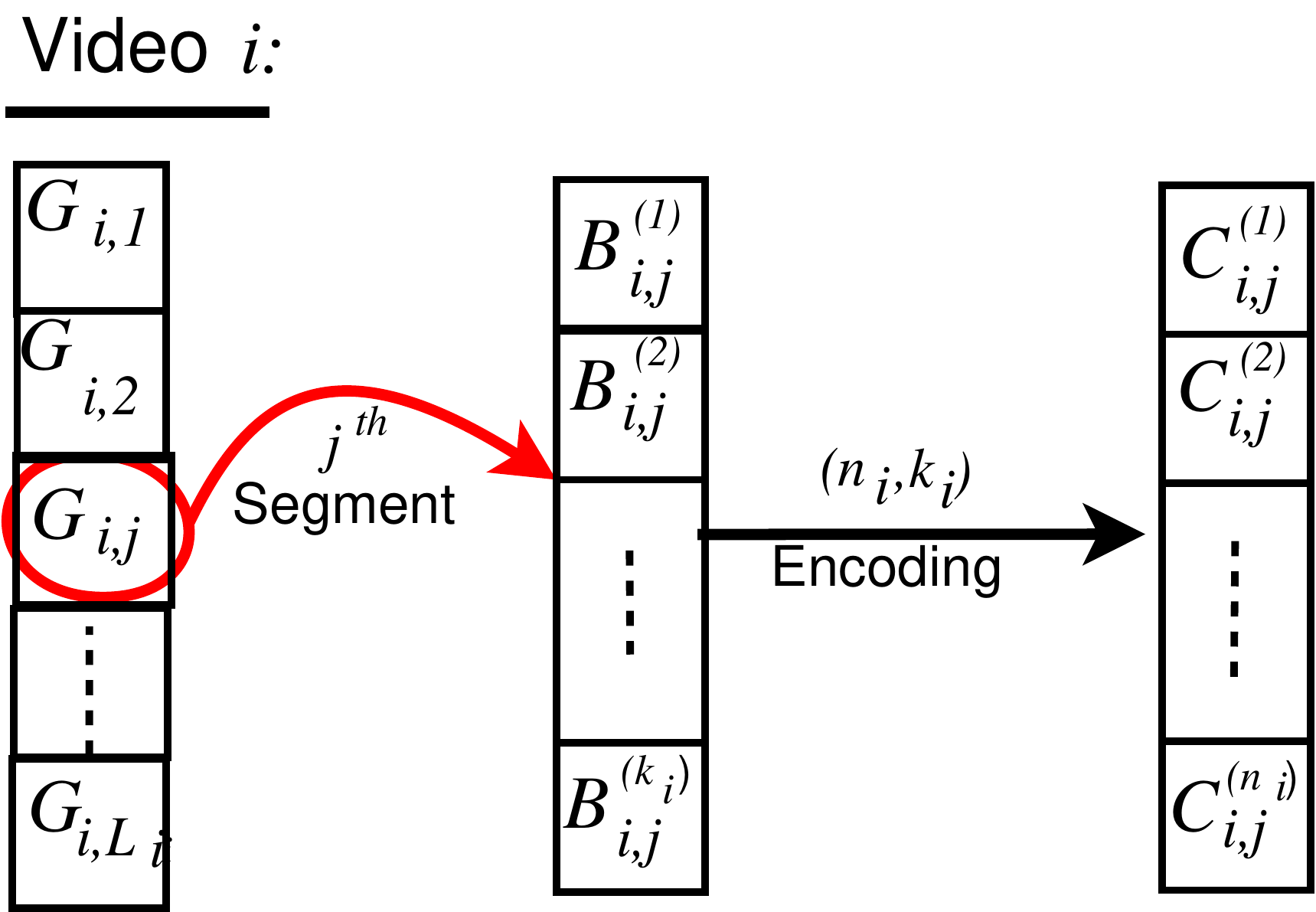}
\caption{A schematic illustrates video fragmentation and erasure-coding processes. Video $i$ is composed of $L_{i}$ segments. Each segments is partitioned into $k_{i}$ chunks and then encoded using an $(n_{i},k_{i})$ MDS code.\label{fig:videoEncoding}}
\end{figure}

We consider a distributed storage system consisting of $m$ heterogeneous servers
(also called storage nodes), denoted by $\mathcal{M}=1,2,...,m$. Each video file $i$, where $i=1,2,...r,$ is divided into $L_{i}$ equal segments, $G_{i,1}, \cdots, G_{i,L_i}$,  each of length $\tau$ sec. Then, each segment $G_{i,j}$ for $j\in\left\{ 1,2,\ldots,L_{i}\right\} $ is  partitioned  into $k_i$ fixed-size chunks  and then
 encoded  using an $(n_i, k_i)$ Maximum Distance Separable (MDS) erasure code to generate $n_i$  distinct chunks for each segment $G_{i,j}$. These coded chunks are denoted as $C_{i,j}^{(1)}, \cdots, C_{i,j}^{(n_i)}$. The encoding setup is illustrated in Figure \ref{fig:videoEncoding}.

\begin{figure}
\includegraphics[scale=0.28,angle=-90]{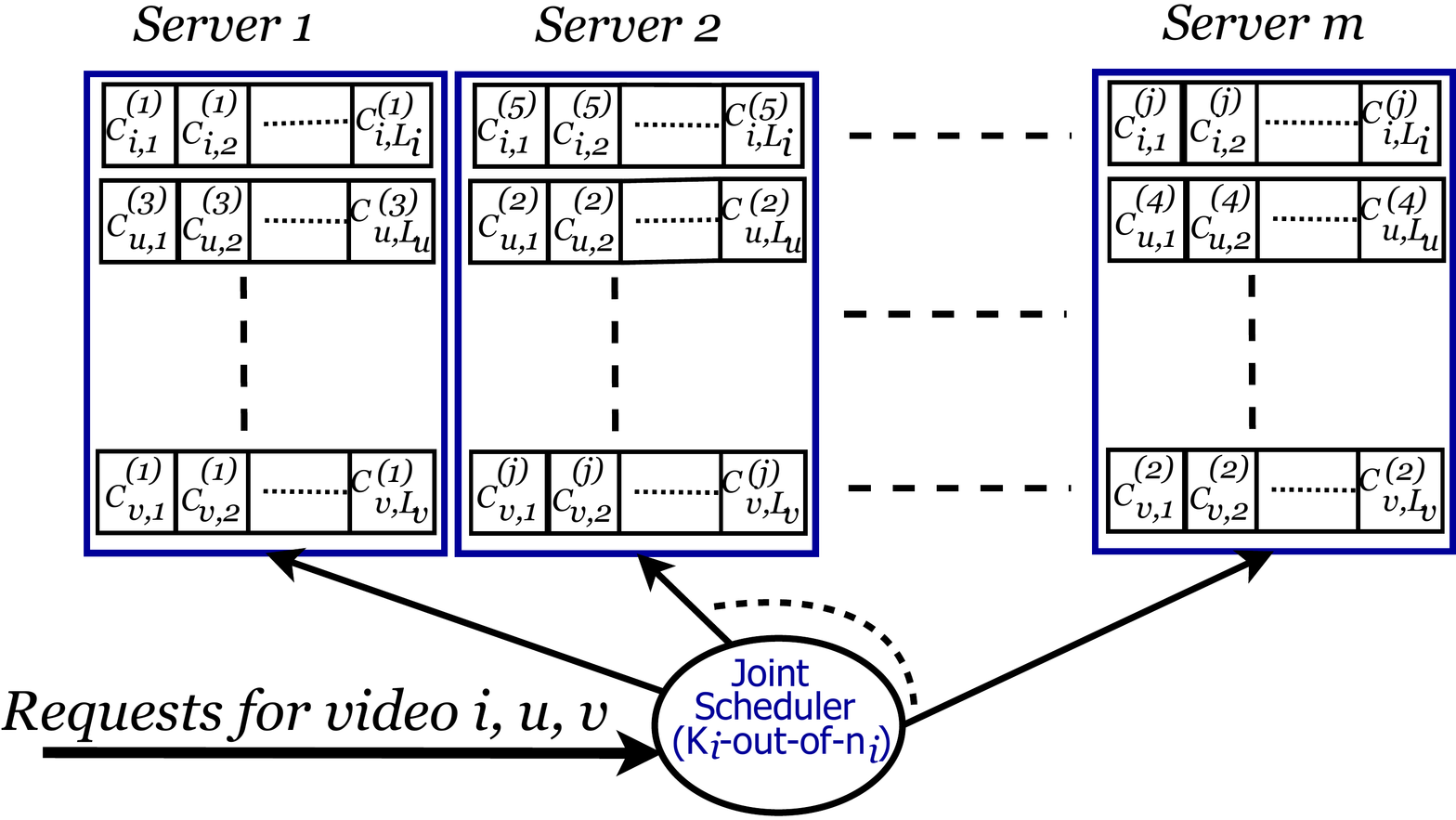}
\vspace{-.3in}
\caption{An Illustration of a distributed storage system
equipped with $m$ nodes and storing $3$ video files assuming $(n_{i},k_{i})$ erasure codes.\label{fig:plcOnServ}}
\end{figure}

The encoded chunks are stored on the disks of $n_i$ distinct storage nodes. These storage nodes are represented by a set $\mathcal{S}_{i}$, such that 
$\mathcal{S}_{i}\subseteq\mathcal{M}$ and $n_{i}=\left|\mathcal{S}_{i}\right|$. Each server $z\in \mathcal{S}_{i}$ stores all the chunks $C_{i,j}^{(g_z)}$ for all $j$ and for some $g_z\in \{1, \cdots, n_i\}$. In other words, each of the $n_i$ storage nodes  stores one of the coded chunks for the entire duration of the video. The placement on the servers is illustrated in Figure \ref{fig:plcOnServ}, where the server $1$ is shown to store first coded chunks of file $i$, third coded chunks of file $u$ and first coded chunks for file $v$.  

The use of $\left(n_{i},k_{i}\right)$ of MDS erasure code introduces a redundancy factor of $n_{i}/k_{i}$ which allows the video to be reconstructed from the video chunks from any subset  of $k_{i}$-out-of-$n_{i}$ servers.  We note that the erasure-code can also help in recovery of the content $i$  as long as $k_i$ of the servers containing file $i$ are available \cite{Dimakis:10}.  Note that replication along $n$ servers is equivalent to choosing $(n,1)$ erasure code.  Hence, when a video $i$ is requested, the request goes to a set $\mathcal{A}_{i}$ of the storage nodes, where  $\mathcal{A}_{i}\subseteq\mathcal{S}_{i}$ and $k_{i}=\left|\mathcal{A}_{i}\right|$. From each server $z \in \mathcal{A}_{i}$, all chunks $C_{i,j}^{(g_z)}$ for all $j$ and the value of $g_z$ corresponding to that placed on server $z$ are requested. The request is illustrated in  Figure \ref{fig:plcOnServ}. In order to play a segment $q$ of video $i$, $C_{i,q}^{(g_z)}$ should have been downloaded from all $z\in \mathcal{A}_{i}$.  We assume that an edge router which is a combination of multiple users is requesting the files. Thus, the connections between the servers and the edge router is considered as the bottleneck. Since the service provider only has control over this part of the network and the last hop may not be under the control of the provider, the service provider can only guarantee the quality-of-service till the edge router. The key used notations are defined in Table \ref{tab:Key-Notations-Used} in Appendix \ref{notation}.

\begin{figure}
\includegraphics[trim=0.2in .5in 0in .3in, clip, width=.45\textwidth]{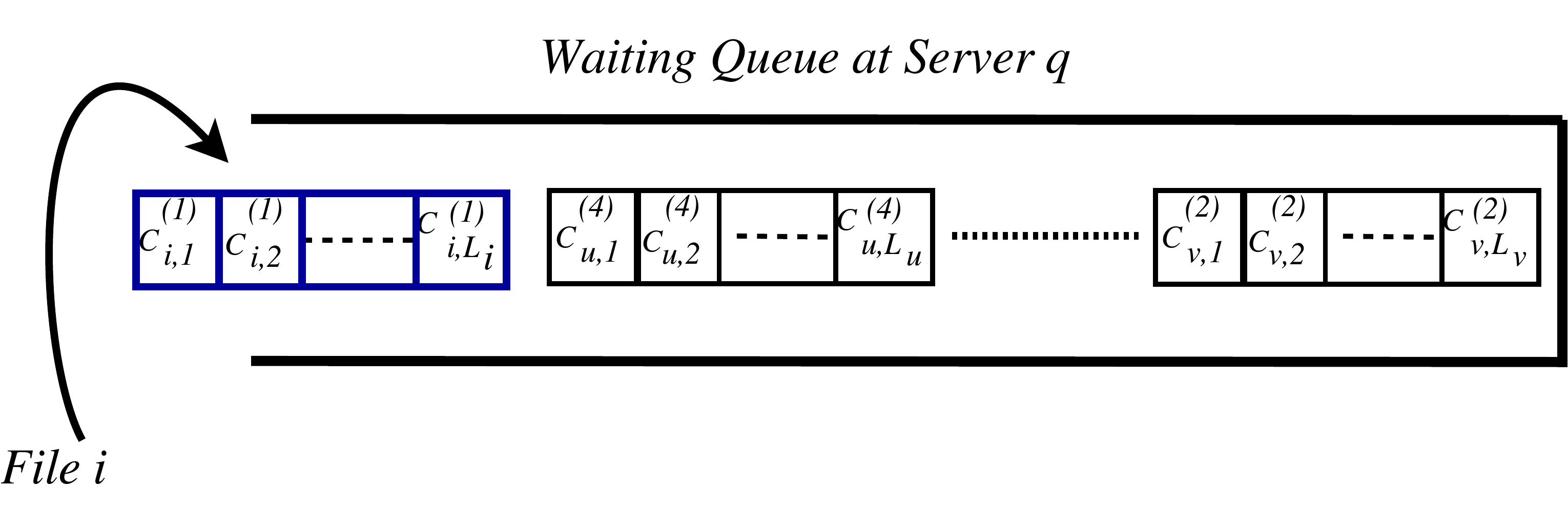}
\caption{An Example of the instantaneous queue status at
server $q$, where $q\in{1,2,...,m}$.\label{fig:sysModel}}
\end{figure}


We assume that the files at each server are served in order of the request in a first-in-first-out (FIFO) policy. Further, the different chunks are processed in order of the duration. This is depicted in  Figure \ref{fig:sysModel}, where for a server $q$, when a file $i$ is requested, all the chunks are placed in the queue where other video requests before this that have not yet been served are waiting. 

In order to schedule the requests for video file $i$ to the $k_i$ servers, the choice of $k_i$-out-of-$n_i$ servers is important. Finding the optimal choice of these servers to compute the latency expressions is an open problem to the best of our knowledge. Thus, this paper uses a policy, called Probabilistic Scheduling, which was proposed in \cite{Xiang:2014:Sigmetrics:2014,Yu_TON}. This policy  allows choice of every possible subset of $k_i$ nodes with certain probability. Upon the arrival of a video file $i$,  we randomly dispatch the batch of $k_{i}$ chunk requests  to appropriate a set of nodes (denoted by
set $\mathcal{A}_{i}$ of servers for file $i$) with predetermined probabilities
($P\left(\mathcal{A}_{i}\right)$ for set $\mathcal{A}_{i}$ and file $i$). Then, each node buffers requests in a local queue and processes in order and independently as explained before. The authors of \cite{Xiang:2014:Sigmetrics:2014,Yu_TON} proved that a probabilistic scheduling policy with feasible probabilities
$\left\{ P\left(\mathcal{A}_{i}\right):\,\forall_{i},\,\mathcal{A}_{i}\right\} $ exists
if and only if there exists conditional probabilities $\pi_{ij}\in\left[0,1\right]$
$\forall i,j$ satisfying

\[
\sum_{j=1}^{m}\pi_{ij}=k_{i}\,\,\,\,\forall i\,\,\,\,\,\,\,\,\,\,\,\,\,\,\,\,\,\,\mbox{and}\,\,\,\,\,\,\pi_{ij}=0\,\,\,\,\,\mbox{if\,\,\,\ensuremath{j\notin \mathcal{S}_{i}}}.
\]

In other words, selecting each node $j$ with probability $\pi_{ij}$ would yield a feasible choice of $\left\{ P\left(\mathcal{A}_{i}\right):\,\forall_{i},\,\mathcal{A}_{i}\right\} $. Thus, we consider the request probabilities $\pi_{ij}$ as the probability that the request for video file $i$ uses server $j$. While the probabilistic scheduling have been used to give bounds on latency of file download, this paper uses the scheduling to give bounds on the QoE for video streaming.

We note that it may not be ideal in practice  for a server to finish one video request before starting another since that increases delay for the future requests. However, this can be easily alleviated by considering that each server has multiple queues (streams) to the edge router which can all be considered as separate servers. These multiple streams can allow multiple parallel videos from the server.  The probabilistic scheduling can choose $k_i$ of the overall queues to access the content. 
%
Possible approaches of extension to accommodate such scenarios are shown in the Appendix \ref{apdx_entend}.



We now describe a queuing model of the distributed storage system.
We assume that the arrival of client requests for each video $i$ form
an independent Poisson process with a known rate $\lambda_{i}$.  The arrival of file requests at node $j$ forms a Poisson Process
with rate $\varLambda_{j}=\sum_{i}\lambda_{i}\pi_{i,j}$ which is
the superposition of $r$ Poisson processes each with rate $\lambda_{i}\pi_{i,j}$.

We assume that the chunk service time for each coded chunk $C_{i,l}^{(g_j)}$ at server $j$,  $X_{j}$,  follows a shifted exponential distribution as has been demonstrated in realistic systems \cite{Yu_TON,CS14}. The service time distribution for the chunk service time at server $j$, $X_{j}$, is given by
the probability distribution function $f_{j}(x)$, which is
\begin{equation}
f_{j}(x)=\begin{cases}
\begin{array}{cc}
\alpha_{j}e^{-\alpha_{j}\left(x-\beta_{j}\right)}\,, & \,\,\,\,\,x\geq\beta_{j}\\
0\,, & \,\,\,\,\,\,x<\beta_{j}
\end{array}\end{cases}.
\end{equation}

We note that exponential distribution is a special case with $\beta_{j}=0$.  We note that the constant delays like the networking delay, and the decoding time can be easily factored into the shift of the shifted exponential distribution. Let $M_{j}(t)=\mathbb{E}\left[e^{tX_{j}}\right]$ be the moment generating function of $X_{j}$.  Then, $M_{j}(t)$ is given as

\begin{equation}
M_{j}(t)=\frac{\alpha_{j}}{\alpha_{j}-t}\,e^{\beta_{j}t}\,\,\,\,\,\,\,\,\,\,t<\alpha_{j} 
\label{M_j_t_1}
\end{equation}

We note that the arrival rates are given in terms of the video files, and the service rate above is provided in terms of the coded chunks at each server. The client plays the video segment after all the $k_i$ chunks for the segment have been downloaded and the previous segment has been played. We also assume that there is a start-up delay of $d_{s}$ (in seconds) for the video which is the duration in which the content can be buffered but not played.  This paper will characterize the stall duration and  stall duration tail probability for this setting. 

\section{Download and Play Times of the Chunks}

In order to understand the stall duration, we need to see the download time of different coded chunks and the play time of the different segments of the video. 

\subsection{Download Times of the Chunks from each Server}
In this subsection, we will quantify the download time of chunk for video file $i$ from server $j$ which has  chunks $C_{i,q}^{(g_j)}$ for all $q = 1, \cdots L_i$. We consider download of $q^{\text{th}}$  chunk $C_{i,q}^{(g_j)}$. As seen in Figure \ref{fig:sysModel}, the download of $C_{i,q}^{(g_j)}$ consists of two components - the waiting time of all the video files in queue before file $i$ request and the service time of all chunks of video file $i$ up to the $q^{\text{th}}$ chunk. Let  $\ensuremath{W_{j}}$ be the random variable corresponding to the waiting time of all the video files in queue before file $i$ request and $Y_{j}^{(q)}$ be the (random) service time of coded chunk $q$ for file $i$ from server $j$. Then, the (random) download time for coded chunk $q\in \{1, \cdots, L_i\}$ for file $i$ at server $j\in \mathcal{A}_{i}$, $D_{i,j}^{(q)}$, is given as 
\begin{equation}
D_{i,j}^{(q)} = W_j + \sum_{v=1}^q Y_{j}^{(v)}. \label{dije}
\end{equation}

We will now find the distribution of $W_j$. We note that this is the waiting time for the video files whose arrival rate is given as $\varLambda_{j}=\sum_{i}\lambda_{i}\pi_{i,j}$. Since the arrival rate of video files is Poisson, the waiting time for the start of video download from a server $j$, $W_j$,  is given by an M/G/1 process.  In order to find the waiting time, we would need to find the service time statistics of the video files. Note that $f_{j}(x)$ gives the service time distribution of only a chunk and not of the video files. 

Video file $i$ consists of $L_{i}$ coded chunks at server $j$ ($j\in \mathcal{S}_{i}$). The total service time for video file $i$ at server $j$ if requested from server $j$, $ST_{i,j} $, is given as  
\begin{equation}
ST_{i,j} = \sum_{v=1}^{L_i} Y_{j}^{(v)}.
\end{equation}

The service time of the video files is given as 
\begin{equation}
R_{j} = \begin{cases}
ST_{i,j}  \quad \text{ with probability } \frac{\pi_{ij}\lambda_{i}}{\Lambda_j} \quad \forall i,
\end{cases}
\end{equation}
since the service time is $ST_{i,j} $ when file $i$ is requested from server $j$. Let $\overline{R}_{j}(s) = {\mathbb E}[e^{-sR_{j} }]$ be the Laplace-Stieltjes Transform of $R_{j}$. 

\begin{lemma}\label{ljlemma}
	The Laplace-Stieltjes Transform of  $R_{j}$, $\overline{R}_{j}(s)=\mathbb{E}\left[e^{-s\overline{R}_{j}}\right]$ is given as
	\begin{equation}
\overline{R}_{j}(s)  = \sum_{i=1}^r \frac{\pi_{ij}\lambda_i}{\Lambda_j}	\left(\frac{\alpha_{j}e^{-\beta_{j}s}}{\alpha_{j}+s}\right)^{L_{i}}\label{eq:servTimeofFile}
	\end{equation}
	\end{lemma}

\begin{proof}
The proof is provided in Appendix \ref{apdx:ljlemma}.
\end{proof}

\begin{corollary}
	The moment generating function for the service time of video files when requested from server $j$, $B_{j}(t)$, is given by
	\begin{equation}
B_{j}(t)  = \sum_{i=1}^r \frac{\pi_{ij}\lambda_i}{\Lambda_j}	\left(\frac{\alpha_{j}e^{\beta_{j}t}}{\alpha_{j}-t}\right)^{L_{i}}\label{eq:servTimeofFileB_j_i}
	\end{equation}
	for any $t>0$, and $t< \alpha_j$.
	\end{corollary}	
\begin{proof}
This corollary follows from (\ref{eq:servTimeofFile})
 by setting $t=-s$.
\end{proof}

The server utilization for the video files at server $j$ is given as $\rho_{j}=\varLambda_{j}\mathbb{E}\left[R_j\right]$. Since $\mathbb{E}\left[R_j\right] = B_j'(0)$, using Lemma \ref{eq:servTimeofFile}, we have







%

\begin{equation}
\rho_{j}=\sum_{i}\pi_{ij}\lambda_{i}L_{i}\left(\beta_{j}+\frac{1}{\alpha_{j}}\right)\label{eq:rho_j}.
\end{equation}


Having characterized the service time distribution of the video files via a Laplace-Stieltjes Transform $\overline{R}_{j}(s) $, the Laplace-Stieltjes Transform of the waiting time $W_{j}$ can be characterized using Pollaczek-Khinchine formula for M/G/1 queues \cite{zwart2000sojourn}, since the request pattern is Poisson and the service time is general distributed. Thus, the Laplace-Stieltjes Transform of the waiting time $W_{j}$ is given as 

\begin{equation}
\mathbb{E}\left[e^{-sW_{j}}\right]=\frac{\left(1-\rho_{j}\right)s\overline{R}_{j}(s)}{s-\Lambda_{j}\left(1-\overline{R}_{j}(s)\right)}\label{eq:E_W_j_laplace}
\end{equation}


Having characterized the Laplace-Stieltjes Transform of the waiting time $W_{j}$ and knowing the distribution of  $Y_{j}^{(v)}$, the Laplace-Stieltjes Transform of the download time $D_{i,j}^{(q)}$ is given as

\begin{equation}
{\mathbb E}[e^{-sD_{i,j}^{(q)}}] = \frac{\left(1-\rho_{j}\right)s\overline{R}_{j}(s)}{s-\Lambda_{j}\left(1-\overline{R}_{j}(s)\right)}\left( \frac{\alpha_{j}}{\alpha_{j}+s}\,e^{-\beta_{j}s}\right)^q.\label{LapOfE_D_ij}
\end{equation}

We note that the expression above holds only in the range of $s$ when  $s-\Lambda_{j}\left(1-\overline{R}_{j}(s)\right)>0$ and $\alpha_{j}+s>0$. Further, the server utilization $\rho_j $ must be less than $1$. The overall download time of all the chunks for the segment $G_{i,q}$ at the client,  $D_{i}^{(q)}$, is given by 

\begin{equation}
D_{i}^{(q)} = \max_{j\in \mathcal{A}_{i}}  D_{i,j}^{(q)}. \label{deq}
\end{equation}

\subsection{Play Time of Each Video Segment}
 Let $T_{i}^{\left(q\right)}$ be the  time at which the segment $G_{i,q}$ is played (started) at the client. The startup delay of the video is $d_s$. Then, the first segment can be played at the maximum of the time the first segment can be downloaded and the startup delay. Thus, 

\begin{eqnarray}
T_{i}^{(1)} & = & \mbox{max }\left(d_{s},\,D_{i}^{(1)}\right).
\label{eq:T_i_qi1}
\end{eqnarray}

For $1<q\le L_i$, the  play time of  segment $q$ of file $i$ is given by the maximum of the time it takes to download the segment and the time at which the previous segment is played plus the time to play a segment ($\tau$ seconds). Thus, the play time of segment $q$ of file $i$, $T_{i}^{(q)}$  can be expressed as 
\begin{eqnarray}
T_{i}^{(q)} & = & \mbox{max }\left(T_{i}^{(q-1)}+\tau,\,D_{i}^{(q)}\right).
\label{eq:T_i_qi}
\end{eqnarray}



Equation \eqref{eq:T_i_qi} gives a recursive equation, which can yield

\begin{eqnarray}
T_{i}^{(L_i)}
 & = &  \mbox{max }\left(T_{i}^{(L_i-1)}+\tau,\,D_{i}^{(L_i)}\right)\nonumber\\
& = &  \mbox{max }\left(T_{i}^{(L_i-2)}+2\tau,\, D_{i}^{(L_i-1)}+\tau,\,D_{i}^{(L_i)}\right)\nonumber\\
& = & \! \mbox{max} \left(d_s+(L_{i}-1)\tau,\right.\nonumber\\
&&\left. \max_{z=2}^{L_i+1}D_{i}^{(z-1)} + (L_i-z+1)\tau \right)\
\label{eq:T_i_qi2}
\end{eqnarray}


Since $D_{i}^{(q)} = \max_{j\in \mathcal{A}_{i}}  D_{i,j}^{(q)}$ from \eqref{deq}, $T_{i}^{(L_{i})}$ can  be written as

\begin{eqnarray}
T_{i}^{(L_{i})}= \max_{z=1}^{L_i+1}\max_{j\in \mathcal{A}_i}\left(p_{i,j,z}\right)\label{eq:T_i_L_i}, \label{eq:pjstart}
\end{eqnarray}where
\begin{eqnarray}
p_{i,j,z}=\begin{cases}
d_{s}+\left(L_{i}-1\right)\tau &,\,\, z=1\\
\\
D_{i,j}^{(z-1)} + (L_i-z+1)\tau \ &,\,\, 2\leq z\leq (L_{i}+1)
\end{cases}\label{eq:pjz}
\end{eqnarray}

We next give the moment generating function of $p_{i,j,z}$ that will be used in the calculations of the QoE metrics in the next sections. Hence, we define the following lemma. 
		
\begin{lemma}\label{lemma_pijz}
	The moment generating function for $p_{i,j,z}$, is given as
	\begin{equation}
\mathbb{E}\left[e^{tp_{i,j,z}}\right]=\begin{cases}
e^{t\left(d_{s}+\left(L_{i}-1\right)\tau\right)} & ,\,z=1\\
e^{t\left(L_{i}+1-z\right)\tau}Z_{i,j}^{(z-1)}\left(t\right) & ,2\leq z\leq L_{i}+1
\end{cases}
\label{eq:momntPjz}
	\end{equation}
where  

\begin{equation}
Z_{i,j}^{(\ell)}\left(t\right) = {\mathbb E}[e^{tD_{i,j}^{(\ell)}}]  = \frac{\left(1-\rho_{j}\right)tB_{j}(t)\left(M_{j}(t)\right)^{\ell}}{t-\Lambda_{j}\left(B_{j}(t)-1\right)}\label{eq:M_D_ij} 
\end{equation}
	\end{lemma}

\begin{proof}
The proof is provided in Appendix \ref{apdx:mgf_pijz}.
\end{proof}

Ideally, the last segment should be completed by time $d_s + L_i \tau$. The difference between $T_{i}^{(L_i)}$ and $d_s + (L_i-1) \tau$ gives the stall duration. Note that the stalls may occur before any segment. This difference will give the sum of durations of all the stall  periods before any segment. Thus, the stall duration for the request of file 
$\delta^{(i)}$ is given as
\begin{equation}
\Gamma^{(i)} = T_{i}^{(L_i)} - d_s - (L_i-1) \tau. \label{eq:base}
\end{equation}
In the next two sections, we will use this stall time to determine the bounds on the mean stall duration and the stall duration tail probability.

\section{Mean Stall Duration}
\label{sec:mean}
In this section, we will provide a bound for the first QoE metric, which is the mean stall duration for a file $i$. We will find the bound by probabilistic scheduling and since probabilistic scheduling is one feasible  strategy, the obtained bound is an upper bound to the optimal strategy. 

Using \eqref{eq:base}, the expected stall time for file $i$ is given as follows
\begin{eqnarray}
\mathbb{E}\left[\Gamma^{(i)}\right] & = & \mathbb{E}\left[T_{i}^{(L_{i})}-d_{s}-\left(L_{i}-1\right)\tau\right]\nonumber \\
\nonumber \\
& = & \mathbb{E}\left[T_{i}^{(L_{i})}\right]-d_{s}-\left(L_{i}-1\right)\tau\label{eq:E_T_s_2}
\end{eqnarray}

 An exact evaluation for the play time of segment $L_{i}$ is hard due to the dependencies between  $p_{jz}$ random variables for different values of $j$ and $z$, where $z\in{(1,2,...,L_{i}+1)}$ and $j\in \mathcal{A}_{i}$. Hence, we derive an upper-bound on the playtime of the segment $L_{i}$ as follows. Using Jensen's inequality \cite{kuczma2009introduction}, we have for $t_i>0$, 
 
 \begin{equation}
e^{t_{i}\mathbb{E}\left[T_{i}^{\left(L_{i}\right)}\right]} \leq \mathbb{E}\left[e^{t_{i}T_{i}^{\left(L_{i}\right)}}\right]. \label{eq:jensen}
 \end{equation}
 
 Thus, finding an upper bound on the moment generating function for $T_i^{(L_i)}$ can lead to an upper bound on the mean stall duration. Thus, we will now bound the moment generating function for $T_i^{(L_i)}$.

\begin{eqnarray}
  \mathbb{E}\left[e^{t_{i}T_{i}^{\left(L_{i}\right)}}\right] & \overset{(a)}{=} & \mathbb{E}\left[\underset{z}{\mbox{max}}  \,\underset{j\in\mathcal{A}_{i}}{\mbox{max}}\,e^{t_{i}p_{ijz}}\right]\nonumber\\
 & = & \mathbb{E}_{\mathcal{A}_{i}}\left[\mathbb{E}\left[\underset{z}{\mbox{max}}\,\underset{j\in{\mathcal{A}_{i}}}{\mbox{max}}\,e^{t_{i}p_{ijz}}|\,\mathcal{A}_{i}\right]\right]\nonumber
\\
 &\overset{(b)}{\leq} & \mathbb{E}_{\mathcal{A}_{i}}\left[\sum_{j\in \mathcal{A}_{i}}\mathbb{E}\left[\underset{z}{\mbox{max}}\,e^{t_{i}p_{ijz}}\right]\right]\nonumber
\\
 & = & \mathbb{E}_{\mathcal{A}_{i}}\left[\sum_{j}F_{ij}\mathbf{1}_{\left\{ j\in\mathcal{A}_{i}\right\} }\right]\nonumber
\\
 & = & \sum_{j}F_{ij}\,\mathbb{E}_{\mathcal{A}_{i}}\left[\mathbf{1}_{\left\{ j\in\mathcal{A}_{i}\right\} }\right]\nonumber
\\
 & = & \sum_{j}F_{ij}\,\mathbb{P}\left(j\in\mathcal{A}_{i}\right)\nonumber
\\
 & \overset{(c)}{=} & \sum_{j}F_{ij}\pi_{ij}
\label{eq:mgf_bound}
\end{eqnarray}
where (a) follows from \eqref{eq:pjstart}, (b) follows by upper bounding $\max_{j\in \mathcal{A}_{i}} $ by $\sum_{j\in \mathcal{A}_{i}} $, (c) follows by probabilistic scheduling where $\mathbb{P}\left(j\in\mathcal{A}_{i}\right) = \pi_{ij}$,   and  $F_{ij}=\mathbb{E}\left[\underset{z}{\mbox{max}}  \, e^{t_{i}p_{ijz}}\right]$. We note that the only inequality here is for replacing the maximum by the sum. Since this term will be inside the logarithm for the mean stall latency, the gap between the term and its bound becomes additive rather than multiplicative.

Substituting \eqref{eq:mgf_bound} in \eqref{eq:jensen}, we have 

\begin{equation}
\mathbb{E}\left[T_{i}^{(L_{i})}\right]\leq\frac{1}{t_{i}}\text{log}\left(\sum_{j=1}^{m}\pi_{ij}F_{ij}\right).\label{eq:ET_i}
\end{equation}

 Let $H_{ij}=\sum_{\ell=1}^{L_{i}}e^{-t_{i}\left(d_{s}+\left(\ell-1\right)\tau\right)}Z_{{i,j}}^{(\ell)}(t_{i})$, where $Z_{{i,j}}^{(\ell)}(t)$ is defined in equation \eqref{eq:M_D_ij}. We  note that $H_{ij}$ can be simplified using the geometric series formula as follows.
 
 \begin{lemma}\label{hijlem}
 	\begin{equation}
 	H_{ij} =  \frac{e^{-t_{i}\left(d_{s}-\tau\right)}\left(1-\rho_{j}\right)t_{i}B_{j}(t_{i})\widetilde{M}_{j}(t_{i})}{t_{i}-\Lambda_{j}\left(B_{j}(t_{i})-1\right)}\frac{1-\left(\widetilde{M}_{j}(t_{i})\right)^{L_{i}}}{\left(1-\widetilde{M}_{j}(t_{i})\right)},
 	\label{eq:H}
 	\end{equation}
 	where $\widetilde{M}_{j}(t_{i})=M_{j}(t_{\text{i}})e^{-t_{i}\tau}$, $M_j(t_i)$ is given in (\ref{M_j_t_1}), and $B_j(t_i)$ is given in (\ref{eq:servTimeofFileB_j_i}). 
 \end{lemma}
 \begin{proof}
 	The proof is provided in Appendix \ref{apdx:hjlem}.
 \end{proof}

 Substituting \eqref{eq:ET_i} in \eqref{eq:E_T_s_2} and some manipulations, the  mean stall duration is bounded as follows.

\begin{theorem}
	The mean stall duration time for file $i$ is bounded by 
	
	\begin{equation}
	\mathbb{E}\left[\Gamma^{(i)}\right]\leq\frac{1}{t_{i}}\text{log}\left(\sum_{j=1}^{m}\pi_{ij}\left(1+H_{ij}\right)\right)\label{eq:T_s_main_stall}
	\end{equation}
	for any $t_{i}>0$, $\rho_{j}=\sum_{i}\pi_{ij}\lambda_{i}L_{i}\left(\beta_{j}+\frac{1}{\alpha_{j}}\right)$, $\rho_{j}<1,\,\text{and }$ \\
	$\,\sum_{f=1}^r\pi_{fj}\lambda_{f}\left(\frac{\alpha_{j}e^{-\beta_{j}t_{i}}}{\alpha_{j}-t_{i}}\right)^{L_{f}}-\left(\Lambda_{j}+t_{i}\right)<0,\,\forall j$.\label{meanthm}
\end{theorem}
\begin{proof}
	The proof is provided in Appendix \ref{apdx:boundmean}.
\end{proof}






Note that Theorem \ref{meanthm} above holds only in the range of $t_i$ when  $t_i-\Lambda_{j}\left(B_{j}(t_i)-1\right)>0$ which reduces to 
$\,\sum_{f=1}^r\pi_{fj}\lambda_{f}\left(\frac{\alpha_{j}e^{-\beta_{j}t_{i}}}{\alpha_{j}-t_{i}}\right)^{L_{f}}-\left(\Lambda_{j}+t_{i}\right)<0,\,\forall i, j$,
and $\alpha_{j}-t_i>0$. Further, the server utilization $\rho_j $ must be less than $1$ for stability of the system.

We note that for the scenario, where the files are downloaded rather than streamed, a metric of interest is the mean download time. This is a special case of our approach when the number of segments of each video is one, or $L_i=1$. Thus, the mean download time of the file follows as a special case of Theorem \ref{meanthm}. We note that the authors of \cite{Xiang:2014:Sigmetrics:2014,Yu_TON} gave an upper bound for mean file download time using probabilistic scheduling. However, the bound in this paper is different since we use moment generating function based bound. The two bounds are compared in Section \ref{sec:num}, and the bounds in this paper are shown to outperform those in \cite{Xiang:2014:Sigmetrics:2014,Yu_TON}.

\section{Stall Duration Tail Probability}\label{sec:tail}
The stall duration tail probability of a file $i$ is defined as the probability that the stall duration tail $\Gamma^{(i)}$ is greater than (or equal) to $x$. Since evaluating  $\text{\text{Pr}}\left(\Gamma^{(i)}\geq x\right)$ in closed-form is hard \cite{MG1:12,Joshi:13,MDS-Queue,Xiang:2014:Sigmetrics:2014,Yu_TON,CS14}, we derive an upper bound on the stall duration tail probability considering Probabilistic Scheduling as follows. 

\begin{align}
\text{\text{Pr}}\left(\Gamma^{(i)}\geq x\right) & \overset{(a)}{=}\text{\text{Pr}}\left(T_{i}^{(L_{i})}\geq x+d_{s}+\left(L_{i}-1\right)\tau\right)\nonumber \\
 & =\text{\text{Pr}}\left(T_{i}^{(L_{i})}\geq\overline{x}\right)\label{eq:Pr_T_s,i}
\end{align}
where $(a)$ follows from (\ref{eq:E_T_s_2}) and $\overline{x}=x+d_{s}+\left(L_{i}-1\right)\tau$. Then,

\begin{equation}
\begin{array}{ccc}
\text{\text{Pr}}\left(T_{i}^{(L_{i})}\geq\overline{x}\right) & \overset{(b)}{=} & \text{\text{Pr}}\left(\underset{z}{\text{max}\,\,}\text{\ensuremath{\underset{j\in\mathcal{A}_{i}}{\text{max}}p_{ijz}}}\geq\overline{x}\right)\\
 & = & \mathbb{E}_{\mathcal{A}_{i},p_{ijz}}\left[\text{\ensuremath{\boldsymbol{1}_{\left(\underset{z}{\text{max}}\,\,\text{\ensuremath{\underset{j\in\mathcal{A}_{i}}{\text{max}}p_{ijz}}}\geq\overline{x}\right)}}}\right]\\
 & \overset{(c)}{=} & \mathbb{E}_{\mathcal{A}_{i},p_{ijz}}\left[\text{\ensuremath{\underset{j\in\mathcal{A}_{i}}{\text{max}}\,\,\boldsymbol{1}_{\left(\underset{z}{\text{max}}\,p_{ijz}\geq\overline{x}\right)}}}\right]\\
 & \overset{(d)}{\leq} & \mathbb{E}_{\mathcal{A}_{i},p_{ijz}}\sum_{j\in\mathcal{A}_{i}}\,\boldsymbol{1}_{\left(\underset{z}{\text{max}}p_{ijz}\geq\overline{x}\right)}\\
 & \overset{(e)}{=} & \sum_{j}\pi_{ij}\mathbb{\mathbb{E}}_{pijz}\left[\,\boldsymbol{1}_{\left(\underset{z}{\text{max}}p_{ijz}\geq\overline{x}\right)}\right]\\
 & = & \sum_{j}\pi_{ij}\mathbb{P}\left(\underset{z}{\text{max}\,\,}p_{ijz}\geq\overline{x}\right)
\end{array}\label{eq:Pr_T_i_L_i_x_bar}
\end{equation}
where $(b)$ follows from (\ref{eq:T_i_L_i}), (c) follows as both max over $z$ and max over $\mathcal{A}_{j}$ are discrete indicies (quantities) and do not depend on other so they can be exchanged, (d) follows by replacing the max by $\sum_{\mathcal{A}_{i}}$, (e) follows from probabilistic scheduling. Using Markov Lemma, we get 
\begin{equation}
\mathbb{P}\left(\underset{z}{\text{max}\,\,}p_{ijz}\geq\overline{x}\right)\leq\frac{\mathbb{E}\left[e^{t_{i}\left(\underset{z}{\text{max}\,\,}p_{ijz}\right)}\right]}{e^{t_{i}\overline{x}}}\label{eq:Mark_lem}
\end{equation}
We further simplify to get 
\begin{align}
\mathbb{P}\left(\underset{z}{\text{max}\,\,}p_{ijz}\geq\overline{x}\right) & \leq\frac{\mathbb{E}\left[e^{t_{i}\left(\underset{z}{\text{max}\,\,}p_{ijz}\right)}\right]}{e^{t_{i}\overline{x}}}\nonumber \\
 & =\frac{\mathbb{E}\left[\underset{z}{\text{max}\,\,}e^{t_{i}p_{ijz}}\right]}{e^{t_{i}\overline{x}}}\nonumber \\
 & \overset{\left(f\right)}{=}\frac{F_{ij}}{e^{t_{i}\overline{x}}}\label{eq:max_z_pjz}
\end{align}
where (f) follows from (\ref{eq:F_ij}). Substituting (\ref{eq:max_z_pjz}) in (\ref{eq:Pr_T_i_L_i_x_bar}), we get the stall duration tail  probability as described in the following theorem (details are provided in Appendix \ref{apdx:boundtail}).

\begin{theorem}
The stall distribution tail probability  for video file $i$ is bounded by 

\begin{equation}
\sum_{j}\frac{\pi_{ij}}{e^{t_{i}x}}\left(1+e^{-t_{i}\left(d_{s}+(L_{i}-1)\tau\right)}\,H_{ij}\right)\
\label{eq:T_stall}
\end{equation}
for any $t_{i}>0$, $\rho_{j}=\sum_{i}\pi_{ij}\lambda_{i}L_{i}\left(\beta_{j}+\frac{1}{\alpha_{j}}\right)$, $\rho_{j}\leq1,$ \\
$\,\sum_{f=1}^r\pi_{fj}\lambda_{f}\left(\frac{\alpha_{j}e^{-\beta_{j}t_{i}}}{\alpha_{j}-t_{i}}\right)^{L_{f}}-\left(\Lambda_{j}+t_{i}\right)<0,\,\forall i,j$, \text{and} $H_{ij}$ is given by (\ref{eq:H}).\label{tailthm}
\end{theorem}


We note that for the scenario, where the files are downloaded rather than streamed, a metric of interest is the latency tail probability which is the probability that the file download latency is greater than $x$.  This is a special case of our approach when the number of segments of each video is one, or $L_i=1$. Thus, the latency tail probability of the file follows as a special case of Theorem \ref{tailthm}. In this special case, the result reduces to that in \cite{Jingxian}.




\section{Optimization Problem Formulation and Proposed Algorithm} \label{sec:probForm}

\subsection{Problem Formulation}

Let $\boldsymbol{\pi} = (\pi_{ij} \forall i=1, \cdots, r \text{ and } j=1, \cdots, m)$, $\boldsymbol{\mathcal{S}}=\left(\mathcal{S}_{1},\mathcal{S}_{2},\ldots,\mathcal{S}_{r}\right)$, and $\boldsymbol{t}=\left(\widetilde{t}_{1},\widetilde{t}_{2},\ldots,\widetilde{t}_{r}; \overline{t}_{1},\overline{t}_{2},\ldots,\overline{t}_{r}\right)$. Note that the values of $t_i$'s used for mean stall duration and the stall duration tail probability can be different and the parameters $\widetilde{t}$ and $\overline{t}$ indicate these parameters for the two cases, respectively. We wish to minimize the two proposed QoE metrics over the choice of scheduling and access decisions. Since this is a multi-objective optimization, the objective can be modeled as a convex combination of the two QoE metrics.

Let $\overline{\lambda}=\sum_{i}\lambda_{i}$ be the total arrival rate. Then, $\lambda_{i}/\overline{\lambda}$ is the ratio of video $i$ requests. The first objective is the minimization of the mean stall duration, averaged over all the file requests, and is given as $\sum_{i}\frac{\lambda_{i}}{\overline{\lambda}}\,\mathbb{E}\left[\Gamma^{\left(i\right)}\right]$. The second objective is the minimization of stall duration tail probability, averaged over all the file requests, and is given as $\sum_{i}\frac{\lambda_{i}}{\overline{\lambda}}\,{\text{Pr}}\left(\Gamma^{(i)}\geq x\right)$. Using the expressions for the mean stall duration and the stall duration tail probability in Sections \ref{sec:mean} and \ref{sec:tail}, respectively, optimization of a convex combination of the two QoE metrics can be formulated as follows.

\begin{align}
\text{min\,\,\,\,\,}\sum_{i}\frac{\lambda_{i}}{\overline{\lambda}}\left[\theta\,\frac{1}{\widetilde{t}_{i}}\text{log}\left(\sum_{j=1}^{m}\pi_{ij}\left(1+\widetilde{H}_{ij}\right)\right)\right.\nonumber \\
+\left.\left(1-\theta\right)\sum_{j}\frac{\pi_{ij}}{e^{\overline{t}_{i}x}}\left(1+e^{-\overline{t}_{i}\left(d_{s}+(L_{i}-1)\tau\right)}\,\overline{H}_{ij}\right)\right]\label{eq:joint_otp_prob}\\
\mbox{s.t.}\,\,\,\,\, \widetilde{H}_{ij}=\frac{e^{-\widetilde{t}_{i}\left(d_{s}-\tau\right)}\left(1-\rho_{j}\right)\widetilde{t}_{i}B_{j}(\widetilde{t}_{i})}{\widetilde{t}_{i}-\Lambda_{j}\left(B_{j}(\widetilde{t}_{i})-1\right)}\widetilde{Q}_{ij}\,\, ,
\label{eq:H_ij}\\
\overline{H}_{ij}=\frac{e^{-\overline{t}_{i}\left(d_{s}-\tau\right)}\left(1-\rho_{j}\right)\overline{t}_{i}B_{j}(\overline{t}_{i})}{\overline{t}_{i}-\Lambda_{j}\left(B_{j}(\overline{t}_{i})-1\right)}\overline{Q}_{ij}\,\, ,
\label{eq:H_ij2}\\
\widetilde{Q}_{ij} =\left[\frac{\widetilde{M}_{j}(\widetilde{t}_{i})\left(1-\left(\widetilde{M}_{j}(\widetilde{t}_{i})\right)^{L_{i}}\right)}{1-\widetilde{M}_{j}(\widetilde{t}_{i})}\right],\,\,
\label{eq:Q_ij}\\
\overline{Q}_{ij} =\left[\frac{\widetilde{M}_{j}(\overline{t}_{i})\left(1-\left(\widetilde{M}_{j}(\overline{t}_{i})\right)^{L_{i}}\right)}{1-\widetilde{M}_{j}(\overline{t}_{i})}\right],\,\,
\label{eq:Q_ij2}\\
\widetilde{M}_{j}({t})=\frac{\alpha_{j}e^{\left(\beta_{j}-\tau\right){t}}}{\alpha_{j}-{t}},\,\, \label{eq:M_telda_opt2}\\ 
{B}_{j}(t)=\sum_{f=1}^r\frac{\lambda_{f}\pi_{fj}}{\Lambda_{j}}\left(\frac{\alpha_{j}e^{\beta_{j}{t}}}{\alpha_{j}-{t}}\right)^{L_f}\, , \label{eq:Bj_const}\\
\widetilde{M}_{j}({t})=\frac{\alpha_{j}e^{\left(\beta_{j}-\tau\right){t}}}{\alpha_{j}-{t}},\,\, \label{eq:M_telda_opt2}\\ 
{B}_{j}(t)=\sum_{f=1}^r\frac{\lambda_{f}\pi_{fj}}{\Lambda_{j}}\left(\frac{\alpha_{j}e^{\beta_{j}{t}}}{\alpha_{j}-{t}}\right)^{L_f}\, , \label{eq:Bj_const}\\ \rho_{j}=\sum_{f=1}^r\pi_{fj}\lambda_{f}L_{f}\left(\beta_{j}+\frac{1}{\alpha_{j}}\right)<1\,\,\,\,\,\,\forall j\label{eq:rho_j}
\end{align}

\begin{eqnarray}
&  & \varLambda_{j}=\sum_{f=1}^r\lambda_{f}\pi_{f,j}\,\,\,\,\,\forall j\label{eq:Lambda_j}\\
&  & \sum_{j=1}^{m}\pi_{i,j}=k_{i}\,\,\,\,\label{eq:sum_ij}\\
&  & \mbox{ \ensuremath{\pi_{i,j}}=0}\,\,\,\mbox{if \ensuremath{j\notin S_{i}}}\,,\ensuremath{\pi_{i,j}}\in\left[0,1\right]\label{eq:pij}\\
&  & \left|\mathcal{S}_{i}\right|=n_{i},\,\,\forall i\label{eq:S_i_and_ni}\\
&  & 0<\widetilde{t}_{i}<\alpha_j,\,\forall j \label{eq:t_i_alpha_j}\\
&  & 0<\overline{t}_{i}<\alpha_j,\,\forall j \label{eq:t_i_alpha_j2}\\
&  & \alpha_{j}\left(e^{(\beta_j-\tau)\widetilde{t}_i}-1\right)+\widetilde{t}_i<0\,,\forall j \label{M_telda_less_1_2} \\
&  & \alpha_{j}\left(e^{(\beta_j-\tau)\overline{t}_i}-1\right)+\overline{t}_i<0\,,\forall j \label{M_telda_less_1} \\
  &  & \sum_{f=1}^r\pi_{fj}\lambda_{f}\left(\frac{\alpha_{j}e^{\beta_{j}\widetilde{t}_{i}}}{\alpha_{j}-\widetilde{t}_{i}}\right)^{L_{f}}-\left(\Lambda_{j}+\widetilde{t}_{i}\right)<0,\,\forall i, j\label{eq:don_pos_cond}\\
  &  & \sum_{f=1}^r\pi_{fj}\lambda_{f}\left(\frac{\alpha_{j}e^{\beta_{j}\overline{t}_{i}}}{\alpha_{j}-\overline{t}_{i}}\right)^{L_{f}}-\left(\Lambda_{j}+\overline{t}_{i}\right)<0,\,\forall i, j\label{eq:don_pos_cond2}\\
&  & \mbox{var.} \ \ \ \  \    \boldsymbol{\pi},\boldsymbol{t}, \mathcal{\boldsymbol{S}}
\label{eq:vars}
\end{eqnarray}

Here, $\theta\in [0,1]$ is a trade-off factor that determines the relative significance of mean and tail probability of the stall durations in the minimization problem. Varying $\theta=0$ to $\theta=1$, the solution for
(\ref{eq:joint_otp_prob}) spans the solutions that  minimize the mean stall duration to ones that minimize the stall duration tail probability. Note that constraint (\ref{eq:rho_j}) gives the load intensity of server $j$. Constraint (\ref{eq:Lambda_j}) gives the aggregate arrival rate $\Lambda_j$ for each node for the  given probabilistic scheduling probabilities $\pi_{ij}$ and arrival rates $\lambda_i$. Constraints \eqref{eq:pij}-\eqref{eq:S_i_and_ni} guarantees that the scheduling probabilities are feasible. Constraints (\ref{eq:t_i_alpha_j})-(\ref{M_telda_less_1})  ensure that $\widetilde{M}_{j}({t})$ exist for each $\widetilde{t}_i$ and $\overline{t}_i$.  Finally, Constraints (\ref{eq:don_pos_cond})-(\ref{eq:don_pos_cond2}) ensure that the moment generating function given in (\ref{eq:M_D_ij}) exists. We note that the the optimization over $\boldsymbol{\pi}$ helps decrease the objective function and gives significant flexibility over choosing the lowest-queue servers for accessing the files. The placement of the video files $\mathcal{\boldsymbol{S}}$ helps separate the highly accessed files on different servers thus reducing the objective. Finally, the optimization over the auxiliary variables  $\boldsymbol{t}$  gives a tighter bound on the objective function. We note that the QoE for file $i$ is weighed by the arrival rate $\lambda_i$ in the formulation. However, general weights can be easily incorporated for weighted fairness or differentiated services.

%


%
%
%
%




Note that the proposed  optimization problem is a mixed integer non-convex  optimization
as we have the placement over $n$ servers and the constraints \eqref{eq:don_pos_cond} and \eqref{eq:don_pos_cond2} are non-convex in $(\boldsymbol{\pi},\boldsymbol{t})$. We also note  the placement may be decided for multiple aggregation VMs simultaneously and may not be a parameter for single aggregation VM. In that case, the proposed algorithm can still be used without an optimization over the placement of video files.  In the next subsection, we will describe the proposed algorithm.

\subsection{Proposed Algorithm}\label{sec:algo}

The joint mean-tail stall duration optimization problem given in  \eqref{eq:joint_otp_prob}-\eqref{eq:vars}  is optimized over three set of variables:  scheduling probabilities $\boldsymbol{\pi}$,  auxiliary parameters $\boldsymbol{t}$, and chunk placement $\boldsymbol{\mathcal{S}}$. Since the problem is non-convex, we propose an iterative algorithm to solve the problem. The proposed algorithm divides the problem into three subproblems that optimize one variable fixing the remaining two. The three sub-problems are labeled as   (i) Access Optimization optimizes  $\boldsymbol{\pi}$ for given $\boldsymbol{\mathcal{S}}$ and $\boldsymbol{t}$,  (ii) Auxiliary Variables Optimization optimizes  $\boldsymbol{t}$ for given $\boldsymbol{\pi}$ and $\boldsymbol{\mathcal{S}}$, and (iii) Placement Optimization optimizes  $\boldsymbol{\mathcal{S}}$ for given $\boldsymbol{\pi}$ and $\boldsymbol{t}$. This algorithm is summarized as follows. 

%
%

\begin{enumerate}[leftmargin=0cm,itemindent=.5cm,labelwidth=\itemindent,labelsep=0cm,align=left]
\item \textbf{Initilization}: Initialize $\boldsymbol{t}$,\,$\boldsymbol{\mathcal{S}}$,
and $\boldsymbol{\pi}$ in the feasible set. 
\item \textbf{While Objective Converge }
\begin{enumerate}

\item  Run Access Optimization using current values of $\boldsymbol{\mathcal{S}}$
and $\boldsymbol{t}$ to get new values of $\boldsymbol{\pi}$
\item  Run Auxiliary Variables Optimization using current values of $\boldsymbol{\mathcal{S}}$
and $\boldsymbol{\pi}$ to get new values of $\boldsymbol{t}$
\item  Run Placement Optimization using current values
of $\boldsymbol{\mathcal{\pi}}$ and $\boldsymbol{t}$ to get new
values of $\boldsymbol{\mathcal{S}}$ and $\boldsymbol{\mathcal{\pi}}$. 
\end{enumerate}
\end{enumerate}

We first initialize $\mathcal{S}_{i}$, $\pi_{ij}$ and $t_i$ $\forall$ $i,j$ such that the choice is feasible for the problem. Then, we do alternating minimization over the three sub-problems defined above. We will describe the three sub-problems along with the proposed solutions for the sub-problems in Appendix \ref{apdx_subprobs}. Each of the three sub-problems are solved by iNner cOnVex
Approximation (NOVA)  algorithm proposed in \cite{scutNOVA}, and is guaranteed to converge to a stationary point.  Since each sub-problem converges (decreasing) and the overall problem is bounded from below, we have the following result.

\begin{theorem} 
	The proposed algorithm  converges to a  stationary point.
\end{theorem}

\vspace{-.3in}

\section{Numerical Results}\label{sec:num}
In this section, we evaluate our proposed  algorithm for optimization of mean and tail probability of stall duration and show the effect of the trade-off of parameter $\theta$. We first study the two extremes where only either mean stall duration objective or tail stall duration probability is considered. Then, we show the tradeoff between the two QoE metrics based on the trade-off parameter $\theta$.


\begin{table}[b]
	\vspace{-.1in}
{\caption{Storage Node Parameters Used in our Simulation (Shift $\beta=10msec$
and rate $\alpha$ in 1/s)}
}

\resizebox{.49\textwidth}{!}{\begin{tabular}{|c|c|c|c|c|c|c|}
\multicolumn{1}{c}{} & \multicolumn{1}{c}{\textbf{Node 1}} & \multicolumn{1}{c}{\textbf{Node 2}} & \multicolumn{1}{c}{\textbf{Node 3}} & \multicolumn{1}{c}{\textbf{Node 4}} & \multicolumn{1}{c}{\textbf{Node 5}} & \multicolumn{1}{c}{\textbf{Node 6}}\tabularnewline
\hline 
{$\alpha_{j}$} & {$18.2298$} & {$24.0552$} & {$11.8750$} & {$17.0526$} & {$26.1912$} & {$23.9059$}\tabularnewline
\hline 
\end{tabular}}

\resizebox{.5\textwidth}{!}{\begin{tabular}{|c|c|c|c|c|c|c|}
\multicolumn{1}{c}{} & \multicolumn{1}{c}{\textbf{Node 7}} & \multicolumn{1}{c}{\textbf{Node 8}} & \multicolumn{1}{c}{\textbf{Node 9}} & \multicolumn{1}{c}{\textbf{Node 10}} & \multicolumn{1}{c}{\textbf{Node 11}} & \multicolumn{1}{c}{\textbf{Node 12}}\tabularnewline
\hline 
{$\alpha_{j}$} & {$27.006$} & {$21.3812$} & {$9.9106$} & {$24.9589$} & {$26.5288$} & {$21.8067$}\tabularnewline
\hline 
\end{tabular}\label{tab:Storage-Nodes-Parameters}
}
\end{table}

\vspace{-.2in}

\subsection{Numerical Setup}

\begin{figure*}[htb]
	\centering
		\hspace{2mm}
		\begin{minipage}{.31\textwidth}
			\centering
			\includegraphics[trim=0.0in 0in 4.2in 0in, clip, width=\textwidth]{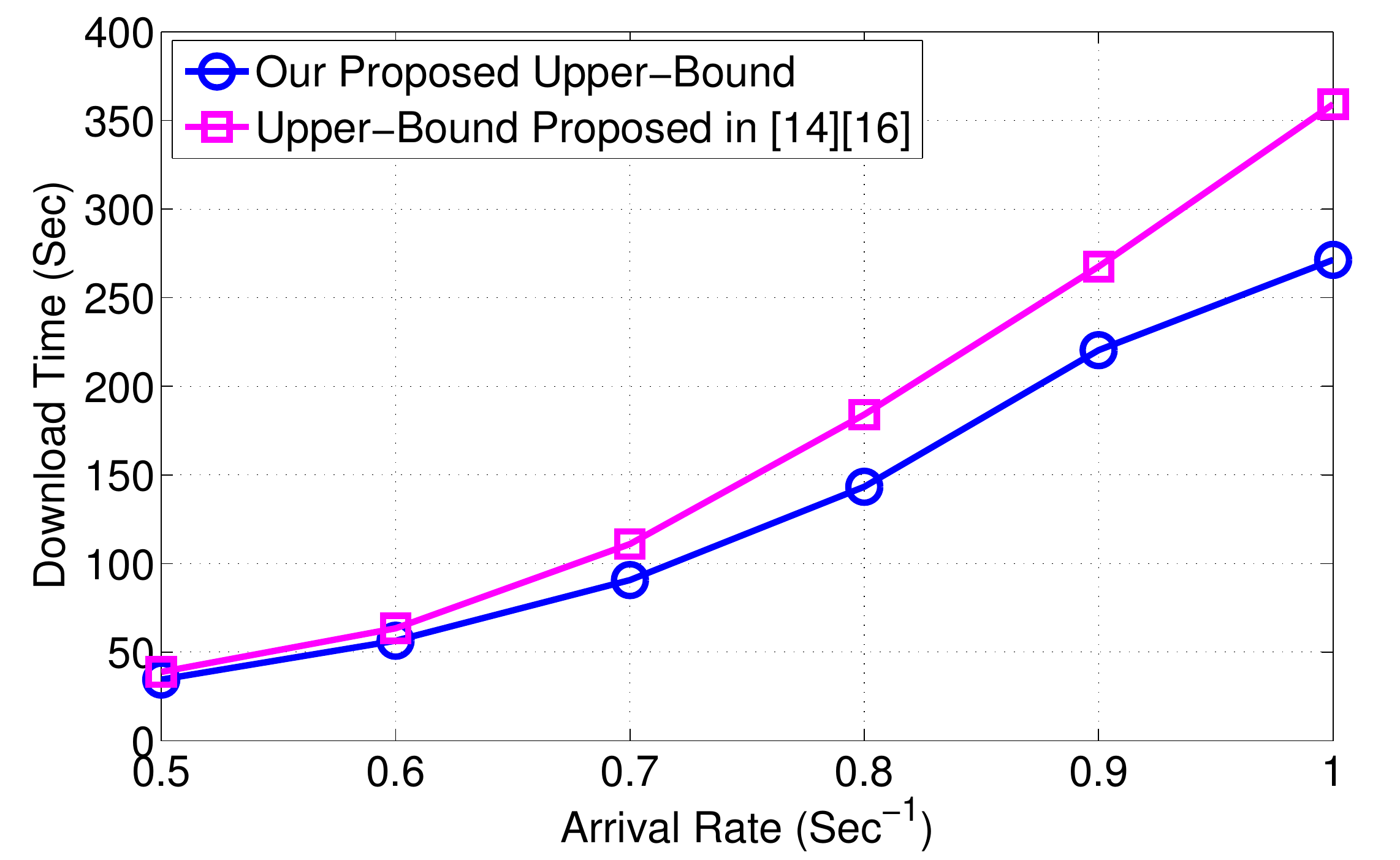}
			\captionof{figure}{Comparison between our upper bound on download time and the upper bound
			proposed in  \cite{Xiang:2014:Sigmetrics:2014,Yu_TON}. 
		}
\label{fig:downloadTime}
		\end{minipage}
	\hspace{1.5mm}
	\begin{minipage}{.31\textwidth}
		\centering
		\includegraphics[trim=0in 0in 0in 0in, clip, width=\textwidth]{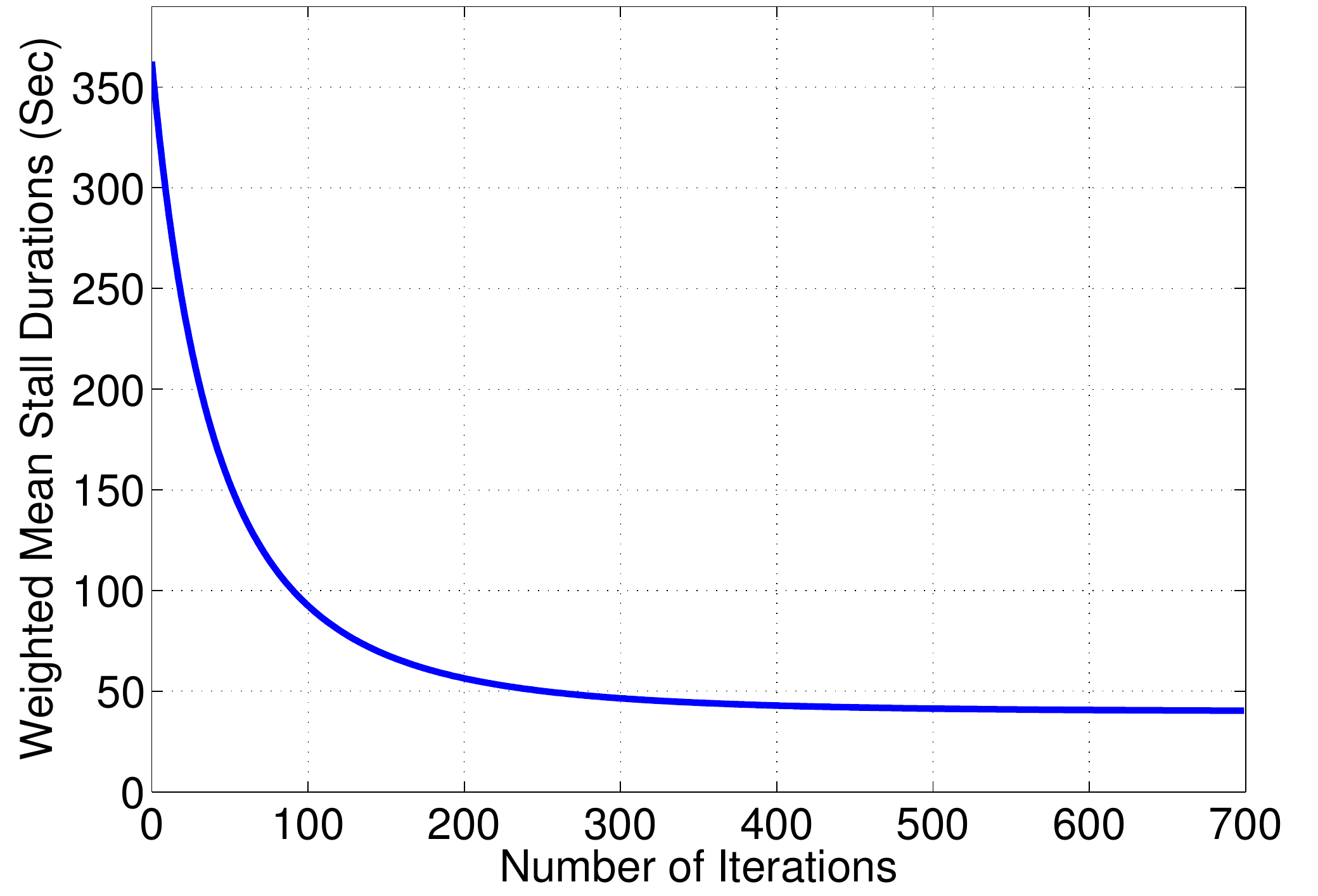}
		\captionof{figure}{Convergence of mean stall duration.}
		\label{fig:ConvgMeanStall}
	\end{minipage}%
	\hspace{1.5mm}
	\begin{minipage}{.31\textwidth}
		\centering
		\includegraphics[trim=0.0in 0in 4.3in 0in, clip,width=\textwidth]{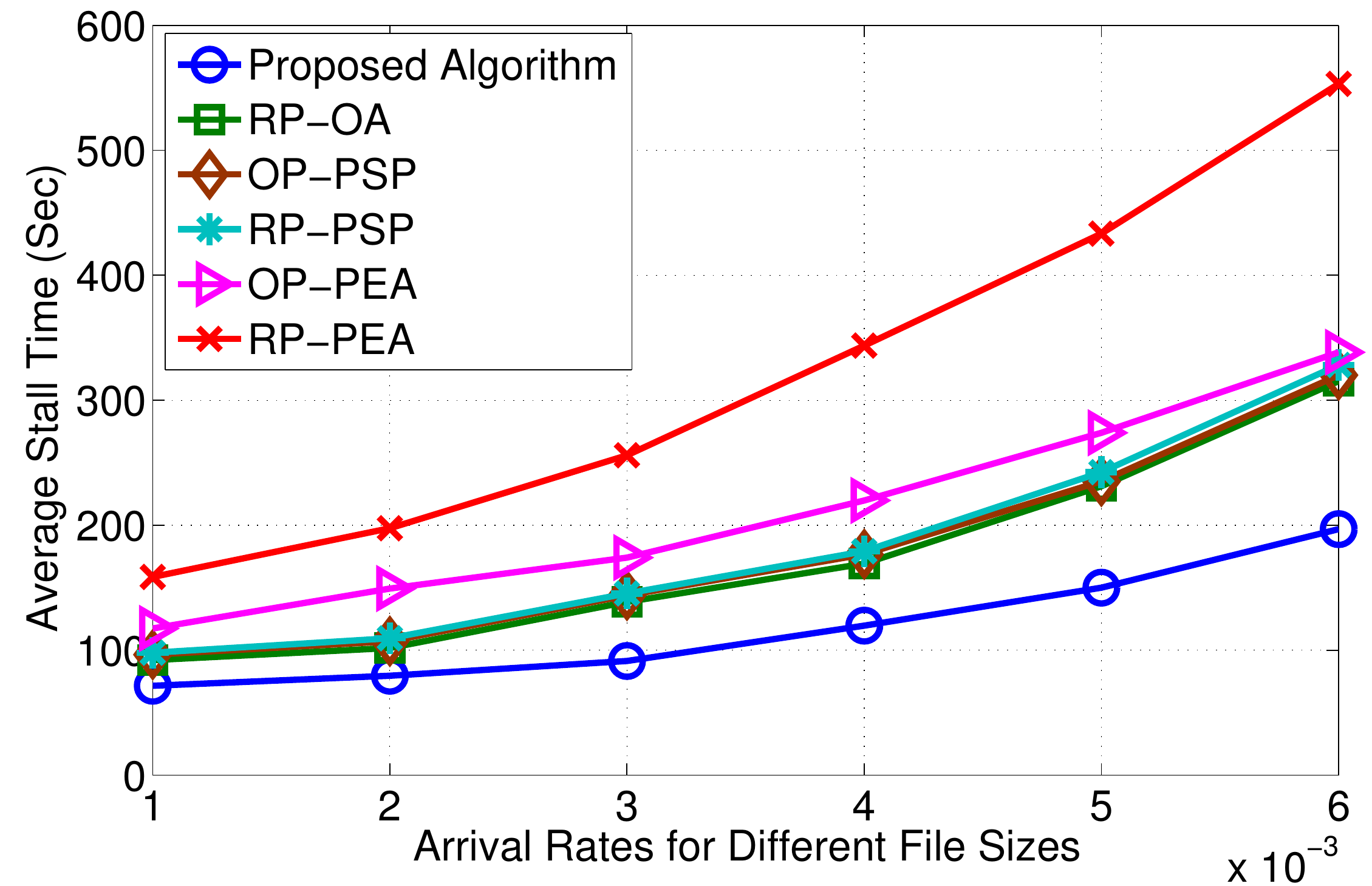}
		\captionof{figure}{Mean stall duration for different video arrival rates with different  video lengths.  }
		\label{fig:meanArrRateDiffSize}
	\end{minipage}
	\vspace{-.1in}
\end{figure*}

 We simulate our algorithm in a distributed storage system of $m=12$ distributed nodes, where each video file uses an $(10,4)$ erasure code.  These parameters were chosen in \cite{Yu_TON} in the experiments using Tahoe testbed.  Further,
$(10,4)$ erasure code is used in HDFS-RAID in Facebook \cite{HDFS_ec} and Microsoft \cite{Asure14}. Unless otherwise explicitly stated, we consider $r=1000$ files, whose sizes are generated based on Pareto distribution \cite{arnold2015pareto} with shape factor of $2$ and scale of $300$, respectively.  We note that the Pareto distribution is considered as it has been widely used in existing
literature  \cite{Vaphase} to
model video files, and file-size distribution over networks.  We also assume that the chunk service time  follows a shifted-exponential distribution with rate $\alpha_{j}$ and shift $\beta_{j}$, whose values are shown in Table I, which are generated at random and kept fixed for the experiments ( Recall that this distribution has been validated in real experiments demonstrated in realistic systems \cite{Yu_TON,CS14}). Unless explicitly stated, the arrival rate for the first $500$ files is $0.002s^{-1}$ while for the next $500$ files is set to be $0.003s^{-1}$. Chunk size $\tau$ is set to be equal to $4$ s. When generating video files, the sizes of the video file sizes are rounded up to the multiple of $4$ sec. We note that a high load scenario is considered for the numerical results. In practice, the load will not be that high. However, higher load helps demonstrate the significant improvement in performance as compared to the lightly loaded scenarios where there are almost no stalls. In order to initialize our algorithm, we use a random placement of files on all the servers. Further, we set $\pi_{ij}=k/n$ on the placed servers with $t_{i}=0.01$ $\forall i$ and $j \in \mathcal{S}_{i}$. However, these choices of $\pi_{ij}$ and $t_{i}$ may not be feasible. Thus, we modify the initialization of  $\boldsymbol{\pi}$  to be closest norm feasible solution  given above values of $\boldsymbol{\mathcal{S}}$ and $\boldsymbol{t}$.  We  compare our proposed approach with five strategies:
 \begin{enumerate}[leftmargin=0cm,itemindent=.5cm,labelwidth=\itemindent,labelsep=0cm,align=left]
\item {\em Random Placement, Optimized Access (RP-OA): }  In this strategy, the placement is chosen at random where any $n$  out of $m$ servers are chosen for each file, where each choice is equally likely. Given the random placement, the variables $\boldsymbol{t}$ and $\boldsymbol{\pi}$ are optimized using the Algorithm in Section \ref{sec:algo}, where $\boldsymbol{\mathcal{S}}$-optimization is not performed.

\item {\em Optimized Placement, Projected Equal Access (OP-PEA): }
The strategy utilizes $\boldsymbol{\pi}$, $\boldsymbol{t}$ and $\boldsymbol{\mathcal{S}}$ as mentioned in the setup. Then, alternating optimization over placement and $\boldsymbol{t}$ are performed using the proposed algorithm.


\item {\em Random Placement, Projected Equal Access (RP-PEA): } In this strategy, the placement is chosen at random where any $n$  out of $m$ servers are chosen for each file, where each choice is equally likely. Further, we set $\pi_{ij}=k/n$ on the placed servers with $t_{i}=0.01$ $\forall i$ and $j \in \mathcal{S}_{i}$. We then modify the initialization of $\boldsymbol{\pi}$ to be closest norm feasible solution  given above values of $\boldsymbol{\mathcal{S}}$ and $\boldsymbol{t}$. Finally, an optimization over $\boldsymbol{t}$ is performed to the objective using Algorithm (\ref{alg:NOVA_Alg1}).

	\item  OP-PSP ({\em  Optimized Placement-Projected Service-Rate Proportional Allocation}) Policy: The joint request scheduler  chooses the access probabilities to be proportional to the service rates of the storage nodes, i.e., $\pi_{ij}=k_{i}\frac{\mu_{j}}{\sum_{j}\mu_{j}}$.  This policy assigns servers proportional to their service rates. These access probabilities are projected toward feasible region  for a uniformly random placed files to ensure stability of the storage system.  With these fixed access probabilities,  the weighted  mean stall duration and stall duration tail probability are optimized over the $\mathbf{t}$, and placement $\boldsymbol{\mathcal{S}}$.

	\item  RP-PSP   ({\em  Random Placement-PSP}) Policy: As compared to the OP-PSP Policy, the chunks are placed uniformly at random. The weighted  mean stall duration and stall duration tail probability are optimized over  the choice of auxiliary variables ${\mathbf t}$.

\subsection{Mean Download Time Comparison}

We note that when the  number of segments, $L_i$, the mean stall duration is the same as the mean download time of the file. Further, the bounds in this paper are different from those given in \cite{Xiang:2014:Sigmetrics:2014,Yu_TON} even though both the works use probabilistic scheduling. We will now  compare our proposed upper-bound on download time of a file with the upper-bound given in \cite{Xiang:2014:Sigmetrics:2014,Yu_TON}. The comparison can be seen in Figure \ref{fig:downloadTime}, where the above service time distributions are used at the servers. We observe that our bound performs better for all values of arrival rate ($\lambda$), and the relative performance increases with the arrival rate. For instance, our bound is $30\%$ lower than that given in \cite{Xiang:2014:Sigmetrics:2014,Yu_TON} when the arrival rate equals $0.8\times\lambda$.



%
\end{enumerate} 
 
\subsection{Mean Stall Duration optimization}
In this subsection, we focus only on minimizing the mean stall duration of all files by setting $\theta=1$, {\em i.e.}, stall duration tail probability is not considered.

\subsubsection*{Convergence of the Proposed Algorithm}
Figure \ref{fig:ConvgMeanStall} shows the convergence  of our proposed algorithm, which alternatively optimizes the mean stall duration of all files over scheduling probabilities $\boldsymbol{\pi}$, auxiliary variables $\boldsymbol{\widetilde{t}}$, and placement $\boldsymbol{\mathcal{S}}$. We notice that for $r=1000$ video files of size 600 sec with $m=12$ storage nodes, the mean stall duration converges to the optimal value within less than $700$ iterations.


\subsubsection*{Effect of Arrival Rate and Video Length} 
\begin{figure}[t]
		\centering
		\includegraphics[trim=0.0in 0in 4.2in 0in, clip, width=0.4\textwidth]{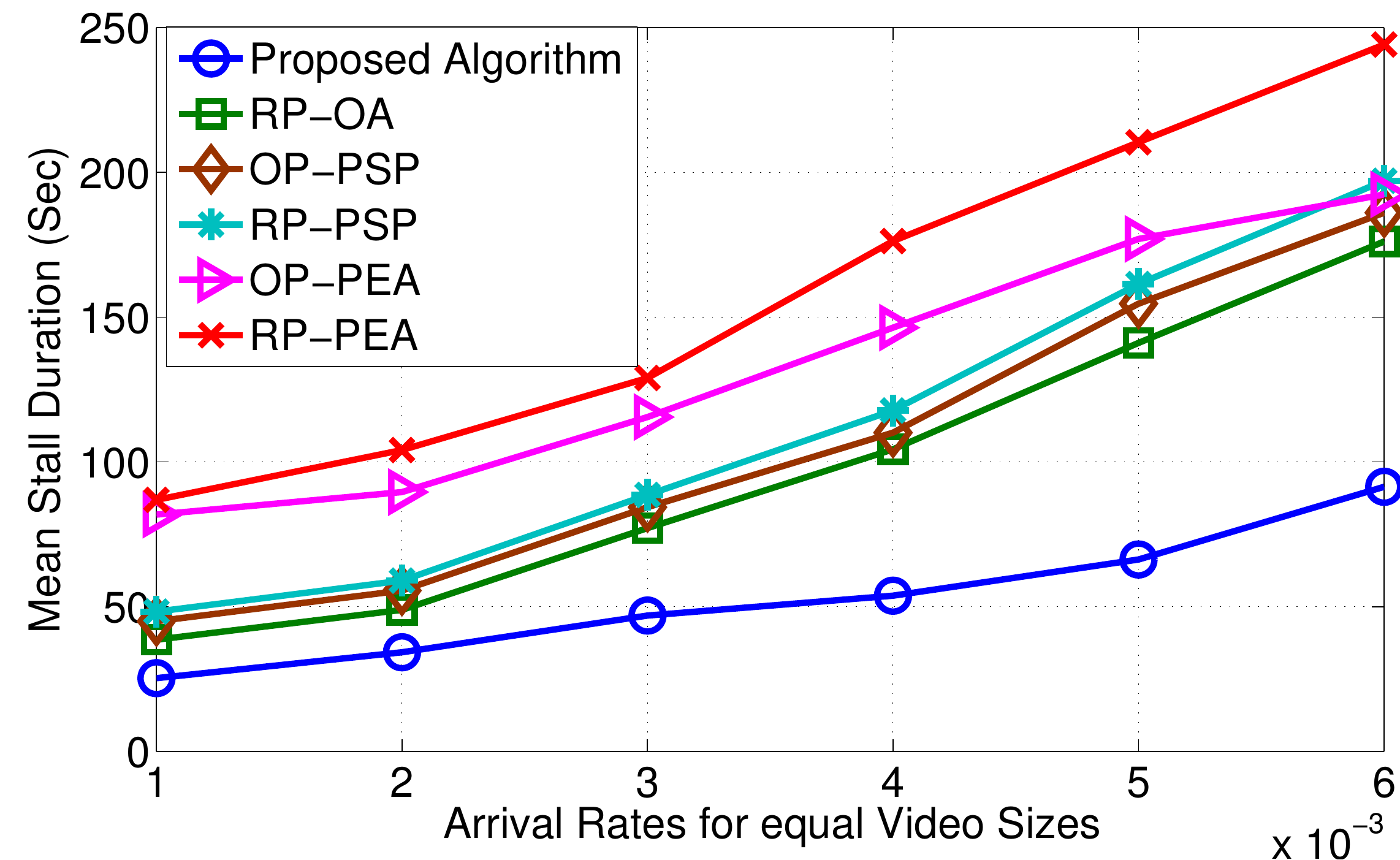}
		\captionof{figure}{Mean stall duration for different video arrival rates for 600s video files. }
		\label{fig:meanArrRateSameSize}
	\end{figure}

Figure
%
%
\ref{fig:meanArrRateDiffSize} shows the effect of different video arrival rates on the mean stall duration for different-size video length.The different size uses the Pareto-distributed lengths described above.  We compare our proposed algorithm with the five baseline  policies and we see that the proposed algorithm outperforms all baseline strategies for the QoE metric of mean stall duration. Thus, both access and placement of files are both important for the reduction of mean stall duration. Further, we see that the mean stall duration increases with arrival rates, as expected. Since the mean stall duration is more significant at high arrival rates, we notice a significant improvement in mean stall duration by about 60\% ( approximately 700s to about 250s) at the highest arrival rate in Figure \ref{fig:meanArrRateDiffSize} as compared to the random placement and projected equal access policy. In Figure
\ref{fig:meanArrRateSameSize}, Appendix \ref{sec:par_enc}, we studied the effect of increasing the arrival rate when the video-sizes are equal with mean of 600 sec. 

\subsection{Stall Duration Tail Probability Optimization}

In this subsection, we consider minimizing the stall duration tail probability, $\mathbb{P}\left(\Gamma^{(i)}\geq x\right)$, by setting $\theta=0$ in (\ref{eq:joint_otp_prob}).

\subsubsection*{Decrease of Stall Duration Tail Probability with $x$}
\begin{figure*}
	\centering
	\begin{minipage}{.31\textwidth}
	\centering
	\includegraphics[trim=0in 0in 4.1in 0in, clip,width=\textwidth]{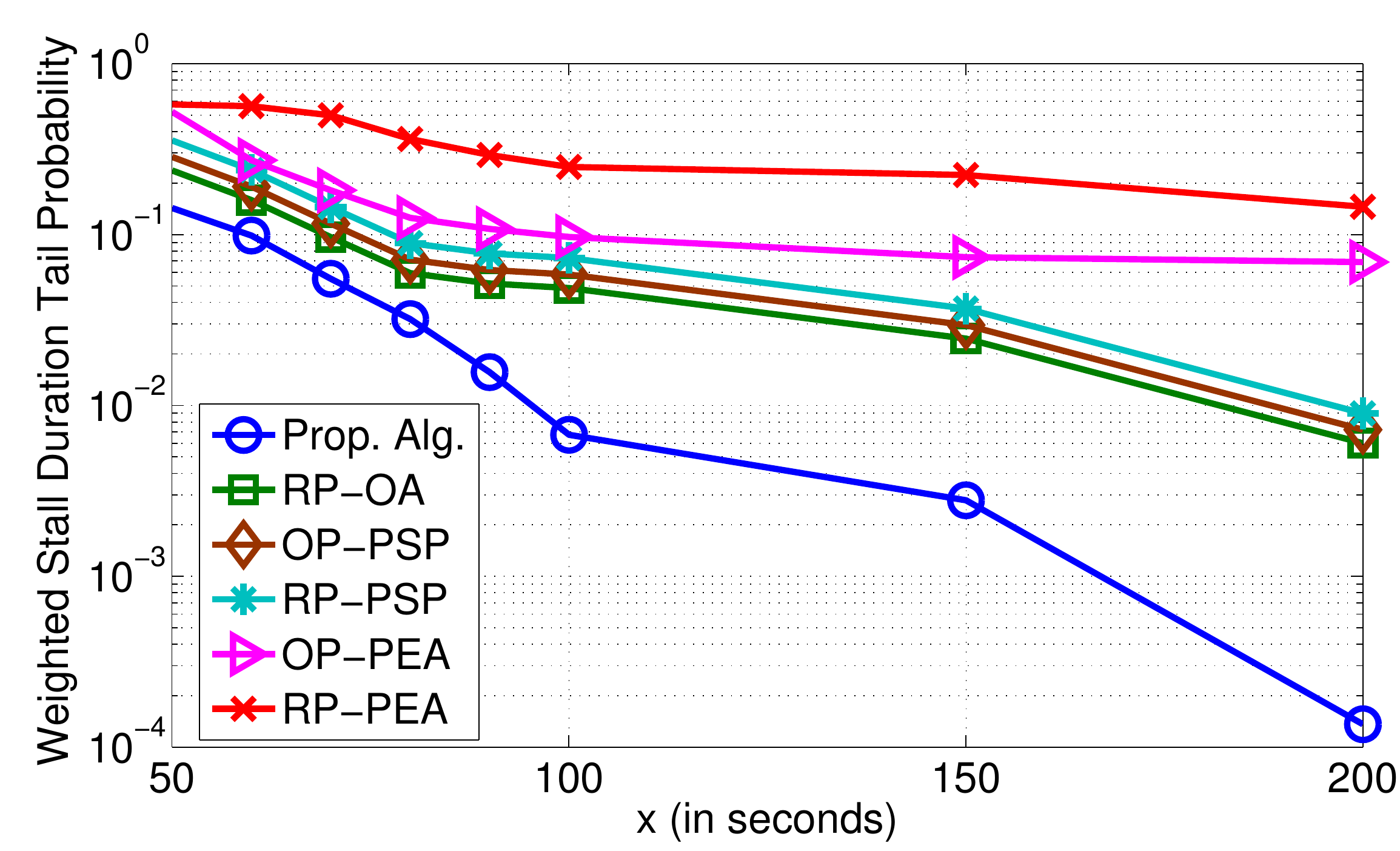}
	\captionof{figure}{Stall duration tail probability for different values of $x$ (in seconds).  }
	\label{fig:tailProbDiffX}
\end{minipage}
		\hspace{2mm}
		\begin{minipage}{.32\textwidth}
			\centering
			\includegraphics[trim=0in 0in 4.3in 0in, clip, width=\textwidth]{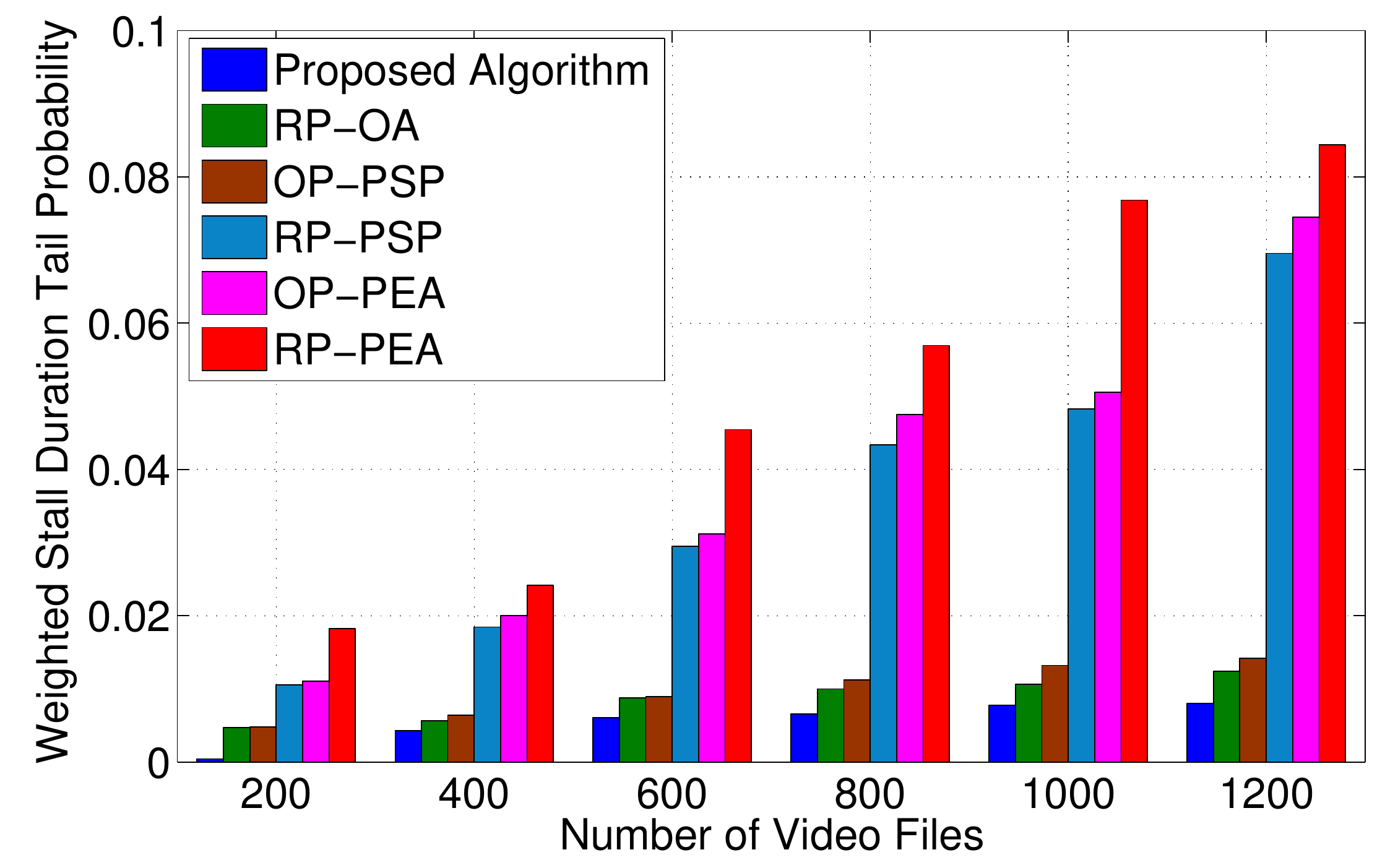}
		\captionof{figure}{Stall duration tail probability for varying number of video files ($x=150$ s).}
		\label{fig:Fig_9_stallLat_vs_NumOfFiles}
		\end{minipage}
		\hspace{2mm}
		\begin{minipage}{.32\textwidth}
			\centering
			\includegraphics[trim=0.4in 0in 0.1in 0in, clip,width=\textwidth]{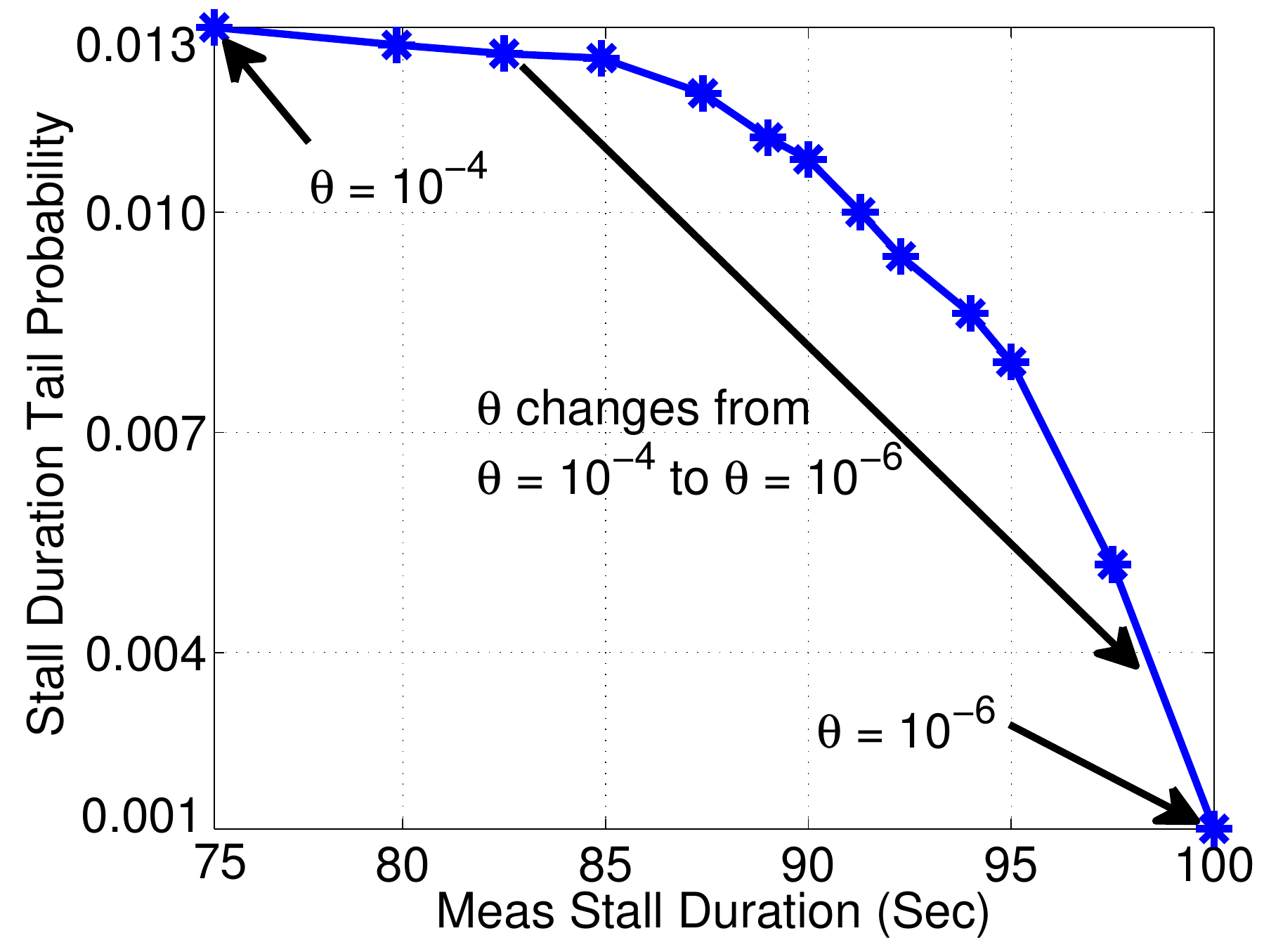}
			\captionof{figure}{Tradeoff between mean stall duration and stall duration tail probability obtained by varying $\theta$.  }
			\label{fig:tradeoff}
		\end{minipage}
	
		\vspace{-.2in}
\end{figure*}

Figure \ref{fig:tailProbDiffX} shows the decay of weighted stall duration tail probability with respect to $x$ (in seconds) for the proposed and the baseline strategies. In order to signify (magnify) the small differences, we plot y-axis in logarithmic scale. We observe that the proposed algorithm gives orders improvement in the stall duration tail probabilities as compared to the baseline strategies. 

\subsubsection*{Effect of the number of video files}

Figure \ref{fig:Fig_9_stallLat_vs_NumOfFiles} demonstrates the effect of increase of the number of video files  ( from $200$ files to $1200$ files whose sizes are defined based on Pareto) on the stall duration tail probability. The stall duration tail probability increases with the number of video files, and the proposed algorithm manages to significantly improve the QoE as compared to the considered baselines. 


\subsection{Tradeoff  between mean stall duration and stall duration tail probability}

If the mean stall duration decreases, intuitively the stall duration tail probability also reduces. Thus, a question arises whether the optimal point for decreasing the mean stall duration and the stall duration tail probability is the same. We answer the question in negative since for $r=1000$ of equal sizes of length 300 sec, we find that at the values of  ($\boldsymbol{\pi}$, $\boldsymbol{\mathcal{S}}$) that optimize the mean stall duration, the stall duration tail probability is 12 times higher as compared to the optimal stall duration tail probability. Similarly, the optimal mean stall duration is 30\% lower as compared to the mean stall duration at the value of ($\boldsymbol{\pi}$, $\boldsymbol{\mathcal{S}}$) that optimizes the stall duration tail probability. Thus, an efficient tradeoff point between the QoE metrics can be chosen based on the point on the curve that is appropriate for the clients.

\vspace{-.2in}
\section{Conclusions}
This paper considers video streaming over cloud where the content is erasure-coded on the distributed servers. Two quality of experience (QoE) metrics related to the stall duration, mean stall duration and stall duration tail probability are characterized with upper bounds. The download and play times of each video segment are characterized to evaluate the QoE metrics. An optimization problem that optimizes the  convex combination of the two QoE metrics  for the choice of placement and access of contents from the servers is formulated. Efficient algorithm is proposed to solve the optimization problem and the numerical results depict the improved performance of the algorithm as compared to the considered baselines.  Some possible future directions are provided in Appendix \ref{apd:future}.


\vspace{-.3in}

\bibliographystyle{IEEEtran}

\bibliography{vidStallRef,allstorage,Tian,ref_Tian2,ref_Tian3,Vaneet_cloud,Tian_rest}
\vspace{-.5in}

\begin{IEEEbiography}[{\includegraphics[width=1in,height=1.25in,clip,keepaspectratio]{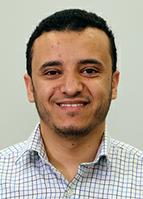}}]{Abubakr O. Al-Abbasi (S'11)} received the B.Sc. and M.Sc. degrees in electronics and electrical communications engineering from Cairo University, Cairo, Egypt, in 2010, and 2014, respectively. He is currently pursuing the Ph.D. degree with Purdue University, USA. From 2011 to 2012, he was a Communications and Networks Engineer with Huawei Company. From 2014 to 2016, he was a Research Assistant with Qatar University, Doha, Qatar. His research interests are in the areas of wireless communications and networking, media streaming, heterogeneous wireless networks, compressive sensing with applications to communications, and signal processing for communications.
\end{IEEEbiography}
\vspace{-.5in}

\begin{IEEEbiography}[{\includegraphics[width=1in,height=1.25in,clip,keepaspectratio]{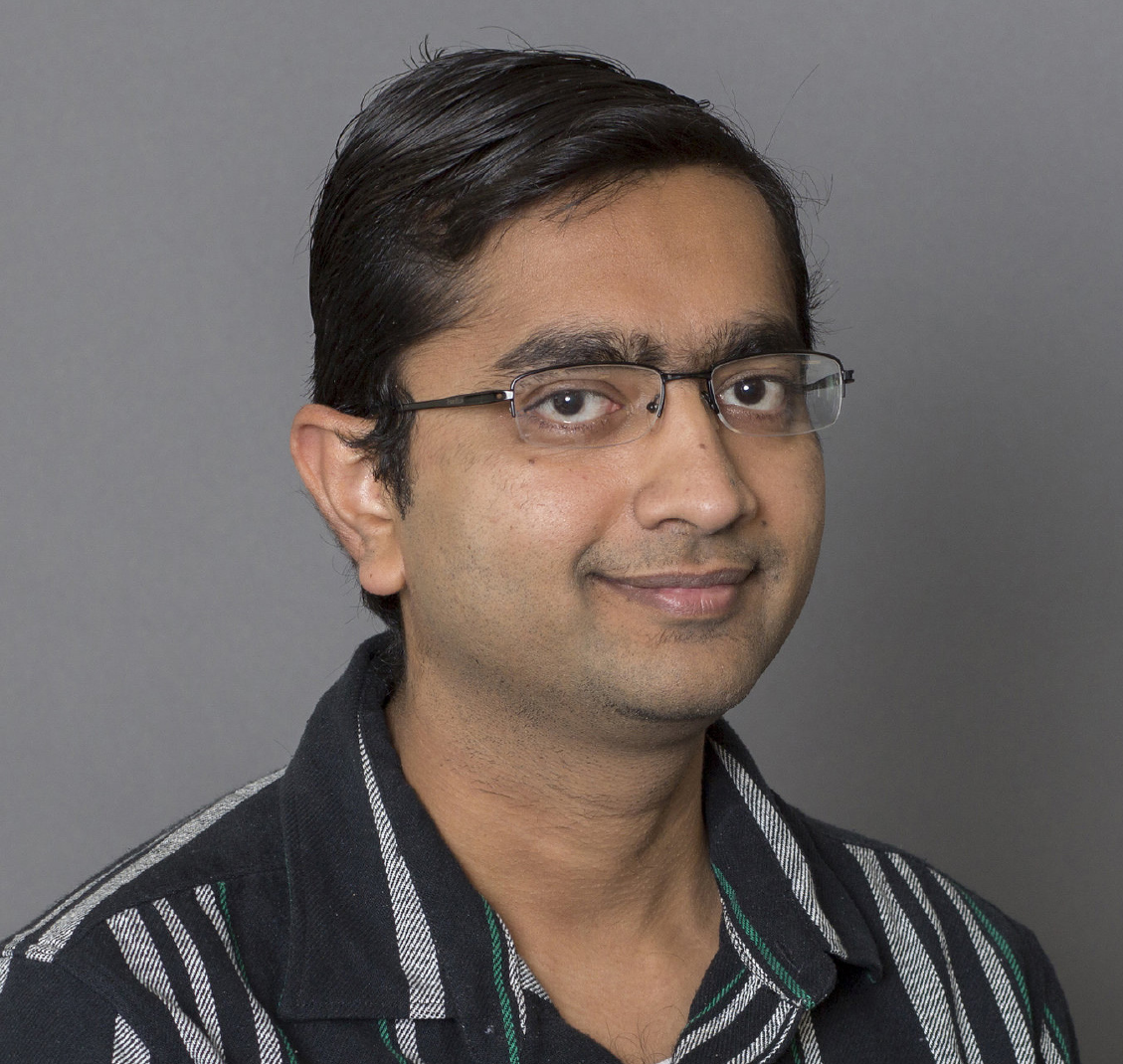}}]{Vaneet Aggarwal (S'08 - M'11 - SM'15)}
	received the B.Tech. degree in 2005 from the Indian Institute of Technology, Kanpur, India, and the M.A. and Ph.D. degrees in 2007 and 2010, respectively from Princeton University, Princeton, NJ, USA, all in Electrical Engineering.
	
	He was a Senior Member of Technical Staff Research at AT\&T Labs-Research, NJ, USA, from 2010 to 2014, and an Adjunct Assistant Professor at Columbia University, NY, USA, from 2013 to 2014. 	He is currently an Assistant Professor at Purdue University, West Lafayette, IN. He is also a VAJRA Adjunct Professor at IISc Bangalore. His current research interests are in communications and networking, video streaming, cloud computing, and machine learning. 
	
Dr. Aggarwal received Princeton University's Porter Ogden Jacobus Honorific Fellowship in 2009, the AT\&T Key Contributor award in 2013, the AT\&T Vice President Excellence Award in 2012, and the AT\&T Senior Vice President Excellence Award in 2014, the 2017 Jack Neubauer Memorial Award recognizing the Best Systems Paper published in the {\em IEEE Transactions on Vehicular Technology}, and the 2018 Infocom Workshop HotPOST Best Paper Award. 	He is on the Editorial Board of the {\it IEEE Transactions on Communications} and the {\it IEEE Transactions on Green Communications and Networking.} 
	
\end{IEEEbiography}


\newpage
\clearpage
\appendices
\section{Key notations used in this paper}\label{notation}

\begin{table}[h]
\caption{Key Notations Used in This Paper\label{tab:Key-Notations-Used}}

\begin{tabular}{|l|>{\raggedright}p{6cm}|}
\hline 
\textbf{Symbol} & \textbf{Meaning}\tabularnewline
\hline 
\hline 
$r$ & Number of video files in system\tabularnewline
\hline 
$m$ & Number of storage nodes \tabularnewline
\hline 
$L_{i}$ & Number of segments for video file $i$\tabularnewline
\hline 
$G_{i,j}$ & Segment $j$ of video file $i$\tabularnewline
\hline 
$\left(n_{i},k_{i}\right)$ & Erasure code parameters for file $i$\tabularnewline
\hline 
$C_{i,j}^{(q)}$ & $q^{th}$ coded chunk of segment $j$ in file $i$ \tabularnewline
\hline 
$\lambda_{i}$ & Possion arrival rate of file $i$\tabularnewline
\hline 
$\pi_{ij}$ & Probability of retrieving chunk of file $i$ from node $j$ using
probabilistic scheduling algorithm\tabularnewline
\hline 
$\mathcal{S}_{i}$ & Set of storage nodes having coded chunks of file $i$\tabularnewline
\hline 
$\mathcal{A}_{i}$ & Set of storage nodes used to access chunks from file $i$\tabularnewline
\hline 
$\left(\alpha_{j},\,\beta_{j}\right)$ & Parameters of Shifted Exponential distribution\tabularnewline
\hline 
$X_{j}$ & Service time distribution of a chunk at node $j$\tabularnewline
\hline 
$M_{j}(t)$ & Moment generating function for the service time of a chunk at node
$j$ $M_{j}(t)=\mathbb{E}\left[e^{tX_{j}}\right]$\tabularnewline
\hline 
$x$ & Parameter indexing stall duration tail probability \tabularnewline
\hline 
$D_{i,j}^{(q)}$ & Download time for coded chunk $q\in\left\{ 1,\ldots,L_{i}\right\} $
of file $i$ from storage node $j$\tabularnewline
\hline 
$R_{j}$ & Service time of the video files \tabularnewline
\hline 
$\overline{R}_{j}$ & Laplace-Stieltjes Transform of $R_{j}$, $\overline{R}_{j}=\mathbb{E}\left[e^{-sR_{j}}\right]$\tabularnewline
\hline 
$B_{j}(t)$ & Moment generating function for the service time of video files $B_{j}(t)=\mathbb{E}\left[e^{tR_{j}}\right]$\tabularnewline
\hline 
$\mu_{j}$ & Mean service time of a chunk from storage node $j$\tabularnewline
\hline 
$\Lambda_{j}$ & Aggregate arrival rate at node $j$\tabularnewline
\hline 
$\rho_{j}$ & Video file request intensity at node $j$\tabularnewline
\hline 
$T_{i}^{(q)}$ & The time at which the segment $G_{i,q}$ is played back\tabularnewline
\hline 
$d_{s}$ & Start-up delay \tabularnewline
\hline 
$\tau$ & Chunk size in seconds\tabularnewline
\hline 
$\Gamma^{(i)}$ & Stall duration tail probability for file $i$\tabularnewline
\hline 
$\theta$ & Trade off factor between mean stall duration and stall duration tail
probability\tabularnewline
\hline 
\end{tabular}
\end{table}
\section{Proof of Lemma \ref{ljlemma}}\label{apdx:ljlemma}

\begin{align}
\overline{R}_{j}(s) & =\sum_{i=1}^r \frac{\pi_{ij}\lambda_{i}}{\Lambda_{j}}\mathbb{E}\left[e^{-s\left(ST_{i,j}\right)}\right]\nonumber \\
& \overset{}{=}\sum_{i=1}^r \frac{\pi_{ij}\lambda_{i}}{\Lambda_{j}}\mathbb{E}\left[e^{-s\left(\sum_{\nu=1}^{L_{i}}Y_{j}^{(\nu)}\right)}\right]\nonumber \\
& =\sum_{i=1}^r \frac{\pi_{ij}\lambda_{i}}{\Lambda_{j}}\left(\mathbb{E}\left[e^{-s\left(Y_{j}^{(1)}\right)}\right]\right)^{L_{i}}\nonumber \\
& =\sum_{i=1}^r \frac{\pi_{ij}\lambda_{i}}{\Lambda_{j}}\left(\frac{\alpha_{j}e^{-\beta_{j}s}}{\alpha_{j}+s}\right)^{L_{i}}
\end{align}

\section{Proof of Lemma \ref{lemma_pijz}\label{apdx:mgf_pijz}}
This follows by substituting
$t=-s$ in (\ref{LapOfE_D_ij}) and $B_{j}(t)$ is given by  (\ref{eq:servTimeofFileB_j_i}) and $M_j(t)$ is given by (\ref{M_j_t_1}). This expressions holds when $t-\Lambda_{j}\left(B_{j}(t)-1\right)>0$ and $t<0  \,\forall j$,  since the moment generating function does not exist if the above does not hold.

\section{Proof of Lemma \ref{hijlem}}\label{apdx:hjlem}
\begin{eqnarray}
&&H_{ij} \nonumber\\
& = & \sum_{\ell=1}^{L_{i}}\left(\frac{e^{-t_{i}\left(d_{s}+\left(\ell-1\right)\tau\right)}\left(1-\rho_{j}\right)t_{i}B_{j}(t_{i})}{t_{i}-\Lambda_{j}\left(B_{j}(t_{i})-1\right)}\left(\frac{\alpha_{j}e^{t_{i}\beta_{j}}}{\alpha_{j}-t_{i}}\right)^{\ell}\right)\nonumber\\
& = & \frac{e^{-t_{i}d_{s}}\left(1-\rho_{j}\right)t_{i}B_{j}(t_{i})}{t_{i}-\Lambda_{j}\left(B_{j}(t_{i})-1\right)}\sum_{\ell=1}^{L_{i}}\left(e^{-t_{i}\left(\ell-1\right)\tau}\left(\frac{\alpha_{j}e^{t_{i}\beta_{j}}}{\alpha_{j}-t_{i}}\right)^{\ell}\right)\nonumber\\
& = & \frac{e^{-t_{i}\left(d_{s}-\tau\right)}\left(1-\rho_{j}\right)t_{i}B_{j}(t_{i})}{t_{i}-\Lambda_{j}\left(B_{j}(t_{i})-1\right)}\sum_{\ell=1}^{L_{i}}\left(e^{-t_{i}\tau}\frac{\alpha_{j}e^{t_{i}\beta_{j}}}{\alpha_{j}-t_{i}}\right)^{\ell}\nonumber\\
& = & \frac{e^{-t_{i}\left(d_{s}-\tau\right)}\left(1-\rho_{j}\right)t_{i}B_{j}(t_{i})}{t_{i}-\Lambda_{j}\left(B_{j}(t_{i})-1\right)}\sum_{\ell=1}^{L_{i}}\left(\frac{\alpha_{j}e^{t_{i}\beta_{j}-t_{i}\tau}}{\alpha_{j}-t_{i}}\right)^{\ell}\nonumber\\
& = & \frac{e^{-t_{i}\left(d_{s}-\tau\right)}\left(1-\rho_{j}\right)t_{i}B_{j}(t_{i})}{t_{i}-\Lambda_{j}\left(B_{j}(t_{i})-1\right)}\times\nonumber\\
&  & \left(M_{j}(t_{i})e^{-t_{i}\tau}\frac{1-\left(M_{j}(t_{i})\right)^{Li}e^{-t_{i}L_{i}\tau}}{1-M_{j}(t_{i})e^{-t_{i}\tau}}\right)\nonumber\\
& = & \frac{e^{-t_{i}\left(d_{s}-\tau\right)}\left(1-\rho_{j}\right)t_{i}B_{j}(t_{i})\widetilde{M}_{j}(t_{i})}{t_{i}-\Lambda_{j}\left(B_{j}(t_{i})-1\right)}\frac{1-\left(\widetilde{M}_{j}(t_{i})\right)^{L_{i}}}{\left(1-\widetilde{M}_{j}(t_{i})\right)}
\end{eqnarray}

\section{Proof of Theorem \ref{meanthm}}\label{apdx:boundmean}

We first find an upper bound on $F_{ij}$   as follows.
\begin{align}
F_{ij} & =\mathbb{E}\left[\underset{z}{\text{max}}\, e^{t_{i}pijz}\right]\nonumber \\
& \overset{(d)}{\leq}\sum_{z}\mathbb{E}\left[e^{t_{i}pijz}\right]\nonumber \\
& \overset{(e)}{=}e^{t_{i}(d_{s}+(L_{i}-1)\tau)}+ \nonumber  \\
& \sum_{z=2}^{L_{i}+1}\frac{e^{t_{i}\left(L_{i}-z+1\right)\tau}\left(1-\rho_{j}\right)t_{i}B_{j}(t_{i})}{t_{i}-\Lambda_{j}\left(B_{j}(t_{i})-1\right)}\left(\frac{\alpha_{j}e^{t_{i}\beta_{j}}}{\alpha_{j}-t_{i}}\right)^{z-1}\nonumber \\
& \overset{(f)}{=}e^{t_{i}(d_{s}+(L_{i}-1)\tau)}+ \nonumber  \\
& \sum_{\ell=1}^{L_{i}}\frac{e^{t_{i}\left(L_{i}-\ell\right)\tau}\left(1-\rho_{j}\right)t_{i}B_{j}(t_{i})}{t_{i}-\Lambda_{j}\left(B_{j}(t_{i})-1\right)}\left(\frac{\alpha_{j}e^{t_{i}\beta_{j}}}{\alpha_{j}-t_{i}}\right)^{\ell}\label{eq:F_ij}
\end{align}
where (d) follows by bounding the maximum by the sum, (e) follows from (\ref{eq:momntPjz}), and (f) follows by substituting $\ell=z-1$.

Further, substituting the bounds \eqref{eq:F_ij} and \eqref{eq:ET_i} in \eqref{eq:E_T_s_2}, the  mean stall duration is bounded as follows.  

\begin{eqnarray}
&&\mathbb{E}\left[\Gamma^{(i)}\right] \nonumber\\
&\leq& \frac{1}{t_{i}}\text{log}\left(\sum_{j=1}^{m}\pi_{ij}\left(e^{t_{i}(d_{s}+(L_{i}-1)\tau)}\right.\right.\nonumber \\
&& \left.\left.+\sum_{\ell=1}^{L_{i}}e^{t_{i}\left(L_{i}-\ell\right)\tau}Z_{{i,j}}^{(\ell)}(t_{i})\right)\right)-\left(d_{s}+\left(L_{i}-1\right)\tau\right)\nonumber \\
&=& \frac{1}{t_{i}}\text{log}\left(\sum_{j=1}^{m}\pi_{ij}\left(e^{t_{i}(d_{s}+(L_{i}-1)\tau)}\right.\right.\nonumber \\
&& \left.\left.+\sum_{\ell=1}^{L_{i}}e^{t_{i}\left(L_{i}-\ell\right)\tau}Z_{{i,j}}^{(\ell)}(t_{i})\right)\right)-\frac{1}{t_{i}}\text{log}\left(e^{t_{i}\left(d_{s}+\left(L_{i}-1\right)\tau\right)}\right)\nonumber \\
&=& \frac{1}{t_{i}}\text{log}\left(\sum_{j=1}^{m}\pi_{ij}\left(1+\sum_{\ell=1}^{L_{i}}e^{-t_{i}\left(d_{s}+\left(\ell-1\right)\tau\right)}Z_{{i,j}}^{(\ell)}(t_{i})\right)\right)\label{eq:ET_s_i_ap}
\end{eqnarray}

\section{Proof of Theorem \ref{tailthm}}\label{apdx:boundtail}

Substituting (\ref{eq:max_z_pjz}) in (\ref{eq:Pr_T_i_L_i_x_bar}), we get 

\begin{eqnarray}
&&\text{\text{Pr}}\left(T_{i}^{(L_{i})}\geq\overline{x}\right) \nonumber\\
& \leq & \sum_{j}\pi_{ij}\mathbb{P}\left(\underset{z}{\text{max}\,\,}p_{ijz}\geq\overline{x}\right)\nonumber\\
& \leq & \sum_{j}\pi_{ij}\frac{F_{ij}}{e^{t_{i}\overline{x}}}\nonumber\\
& \overset{(g)}{\leq} & \sum_{j}\frac{\pi_{ij}}{e^{t_{i}\overline{x}}}\left(e^{t_{i}(d_{s}+(L_{i}-1)\tau)}+H_{ij}\right)\nonumber\\
& = & \sum_{j}\frac{\pi_{ij}}{e^{t_{i}\left(x+d_{s}+(L_{i}-1)\tau\right)}}\left(e^{t_{i}(d_{s}+(L_{i}-1)\tau)}+H_{ij}\right)\nonumber\\
& = & \sum_{j}\frac{\pi_{ij}}{e^{t_{i}x}}\left(1+e^{-t_{i}\left(d_{s}+(L_{i}-1)\tau\right)}\,H_{ij}\right)
\label{eq:Pr_T_i_L_i_x_bar_final}
\end{eqnarray}
where (g) follows from (\ref{eq:F_ij}) and $H_{ij}$ is given by (\ref{eq:H}).

\section{Description of the Algorithms for the Three Sub-Problems} \label{apdx_subprobs}

\subsection{Access Optimization}

Given the placement and the auxiliary variables, this subproblem can be written as follows. 

\textbf{Input: $\boldsymbol{t}$, $\mathcal{\boldsymbol{S}}$ }

\textbf{Objective:} $\qquad\quad\;\;$min $\left(\ref{eq:joint_otp_prob}\right)$

$\hphantom{\boldsymbol{\text{Objective:}\,}}\qquad\qquad$s.t. \eqref{eq:rho_j}, \eqref{eq:Lambda_j}, \eqref{eq:sum_ij}, \eqref{eq:pij},  \eqref{eq:don_pos_cond},  \eqref{eq:don_pos_cond2}

$\hphantom{\boldsymbol{\text{Objective:}\,}}\qquad\qquad$var. $\boldsymbol{\pi}$


In order to solve this problem, we have used iNner cOnVex
Approximation (NOVA)  algorithm proposed in \cite{scutNOVA} to solve this sub-problem. The key idea for this algorithm is that the non-convex objective function is replaced by suitable convex approximations at which convergence to a stationary solution of the original non-convex optimization is established. NOVA solves the approximated function efficiently and maintains feasibility in each iteration. The objective function can be approximated by a convex one (e.g., proximal gradient-like approximation) such that the first order properties are preserved \cite{scutNOVA},  and this convex approximation can be used in NOVA algorithm.

Let $\widetilde{U}\left(\boldsymbol{\pi};\boldsymbol{\pi^\nu}\right)$ be the
convex approximation at iterate $\boldsymbol{\pi^\nu}$ to the original non-convex problem $U\left(\boldsymbol{\pi}\right)$, where $U\left(\boldsymbol{\pi}\right)$ is given by (\ref{eq:joint_otp_prob}). Then, a valid choice of  $\widetilde{U}\left(\boldsymbol{\pi};\boldsymbol{\pi^\nu}\right)$ is the first order approximation of  $U\left(\boldsymbol{\pi}\right)$, e.g., (proximal) gradient-like approximation, i.e.,  
\begin{equation}
\widetilde{U}\left(\boldsymbol{\pi},\boldsymbol{\pi^\nu}\right)=\nabla_{\boldsymbol{\pi}}U\left(\boldsymbol{\pi^\nu}\right)^{T}\left(\boldsymbol{\pi}-\boldsymbol{\pi^\nu}\right)+\frac{\tau_{u}}{2}\left\Vert \boldsymbol{\pi}-\boldsymbol{\pi^\nu}\right\Vert ^{2},\label{eq:U_x_u_bar}
\end{equation}
where $\tau_u$ is a regularization parameter. Note that all the constraints \eqref{eq:rho_j}, \eqref{eq:Lambda_j}, \eqref{eq:sum_ij},  \eqref{eq:pij}, \eqref{eq:don_pos_cond}, and \eqref{eq:don_pos_cond2} are linear in $\boldsymbol{\pi_{i,j}}$.  The NOVA Algorithm for optimizing $\boldsymbol{\pi}$ is described in  Algorithm \ref{alg:NOVA_Alg1Pi}. Using the convex approximation $\widetilde{U}\left(\boldsymbol{\pi};\boldsymbol{\pi^\nu}\right)$, the minimization steps in Algorithm \ref{alg:NOVA_Alg1Pi} are convex, with linear constraints and thus can be solved using a projected gradient descent algorithm. A step-size ($\gamma
$)  is also used in the update of the iterate $\boldsymbol{\pi}^{\nu}$. Note that the iterates $\left\{ \boldsymbol{\pi}^{(\nu)}\right\} $ generated by the algorithm are all feasible for the original problem and, further, convergence is guaranteed, as shown in \cite{scutNOVA} and described in the following lemma. 

\begin{lemma}  \label{lem_pi}
	For fixed placement $\boldsymbol{\mathcal{S}}$ and $\boldsymbol{t}$, 
	the optimization of our problem over 
	$\boldsymbol{\pi}$ generates a sequence of  decreasing
	objective values and therefore is guaranteed to converge to a stationary point.
\end{lemma}

\begin{algorithm}[h]
	\caption{NOVA Algorithm to solve Access Optimization sub-problem\label{alg:NOVA_Alg1Pi}}
	
	\begin{enumerate}
		\item \textbf{Initialize} $\nu=0$, $k=0$,$\gamma^{\nu}\in\left(0,1\right]$,
		$\epsilon>0$,$\boldsymbol{\pi}^{0}$ such that $\boldsymbol{\pi}^{0}$
		is feasible ,
		\item \textbf{while} $\mbox{obj}\left(k\right)-\mbox{obj}\left(k-1\right)\geq\epsilon$
		\item $\quad$//\textit{\small{}Solve for $\boldsymbol{\pi}^{\nu+1}$ with
			given $\boldsymbol{\pi}^{\nu}$}{\small \par}
		\item $\quad$\textbf{Step 1}: Compute $\boldsymbol{\widehat{\pi}}\left(\boldsymbol{\pi}^{\nu}\right),$
		the solution of $\boldsymbol{\widehat{\pi}}\left(\boldsymbol{\pi}^{\nu}\right)=$$\underset{\boldsymbol{\pi}}{\text{argmin}}$
		$\boldsymbol{\widetilde{U}}\left(\boldsymbol{\pi},\boldsymbol{\pi}^{\nu}\right)\,\, $ s.t.
		$\left(\ref{eq:rho_j}\right)$, $\left(\ref{eq:Lambda_j}\right)$, $\left(\ref{eq:sum_ij}\right)$, $\left(\ref{eq:pij}\right)$, $(\ref{eq:t_i_alpha_j})$,  $\left(\ref{eq:don_pos_cond}\right)$, solved using  projected gradient descent 
		
		\item $\quad$\textbf{Step 2}: $\ensuremath{\boldsymbol{\pi}^{\nu+1}=\boldsymbol{\pi}^{\nu}+\gamma^{\nu}\left(\widehat{\boldsymbol{\pi}}\left(\boldsymbol{\pi}^{\nu}\right)-\boldsymbol{\pi}^{\nu}\right)}$.
		\item $\quad$//\textit{\small{}update index}{\small \par}
		\item \textbf{Set} $\ensuremath{\nu\leftarrow\nu+1}$
		\item \textbf{end while}
		\item \textbf{output: }$\ensuremath{\widehat{\boldsymbol{\pi}}\left(\boldsymbol{\pi}^{\nu}\right)}$
	\end{enumerate}
\end{algorithm}


\begin{algorithm}[h]
	\caption{NOVA Algorithm to solve Auxiliary Variables  Optimization sub-problem\label{alg:NOVA_Alg1}}
	
	\begin{enumerate}
		\item \textbf{Initialize} $\nu=0$,$\gamma^{\nu}\in\left(0,1\right]$, $\epsilon>0$, $\boldsymbol{t}^{0}$ 
		such that $\boldsymbol{t}^{0}$  is feasible,
		\item \textbf{while} $\mbox{obj}\left(\nu\right)-\mbox{obj}\left(\nu-1\right)\geq\epsilon$
		\item $\quad$//\textit{\small{}Solve for  $\boldsymbol{t}^{\nu+1}$ with
			given $\boldsymbol{t}^{\nu}$}{\small \par}
		\item $\quad$\textbf{Step 1}: Compute $\boldsymbol{\widehat{t}}\left(\boldsymbol{t}^{\nu}\right),$
		the solution of  $\boldsymbol{\widehat{t}}\left(\boldsymbol{t}^{\nu}\right)=$$\underset{\boldsymbol{t}}{\text{argmin}}$
		$\boldsymbol{\overline{U}}\left(\boldsymbol{t},\boldsymbol{t}^{\nu}\right)$, s.t. \eqref{eq:t_i_alpha_j}, \eqref{M_telda_less_1}, and  \eqref{eq:don_pos_cond}
		using projected gradient descent
		\item $\quad$\textbf{Step 2}: $\ensuremath{\boldsymbol{t}^{\nu+1}=\boldsymbol{t}^{\nu}+\gamma^{\nu}\left(\widehat{\boldsymbol{t}}\left(\boldsymbol{t}^{\nu}\right)-\boldsymbol{t}^{\nu}\right)}$.
		\item $\quad$//\textit{\small{}update index}{\small \par}
		\item \textbf{Set} $\ensuremath{\nu\leftarrow\nu+1}$
		\item \textbf{end while}
		\item \textbf{output: }$\ensuremath{\widehat{\boldsymbol{t}}\left(\boldsymbol{t}^{\nu}\right)}$
	\end{enumerate}
\end{algorithm}

\subsection{Auxiliary Variables Optimization }
Given the placement and the access variables, this subproblem can be written as follows. 

\textbf{Input: $\boldsymbol{\pi}$, $\mathcal{\boldsymbol{S}}$ }

\textbf{Objective:} $\qquad\quad\;\;$min $\left(\ref{eq:joint_otp_prob}\right)$

$\hphantom{\boldsymbol{\text{Objective:}\,}}\qquad\qquad$s.t. \eqref{eq:t_i_alpha_j}, \eqref{eq:t_i_alpha_j2}, \eqref{M_telda_less_1_2},\eqref{M_telda_less_1}, \eqref{eq:don_pos_cond},
\eqref{eq:don_pos_cond2},

$\hphantom{\boldsymbol{\text{Objective:}\,}}\qquad\qquad$var. $\boldsymbol{t}$

Similar to Access Optimization, this optimization can be solved using NOVA algorithm. The constraints \eqref{eq:t_i_alpha_j} and \eqref{eq:t_i_alpha_j2} are linear in $\boldsymbol{t}$. The next two Lemmas show that the constraints \eqref{M_telda_less_1_2},  \eqref{M_telda_less_1}, \eqref{eq:don_pos_cond}, and  \eqref{eq:don_pos_cond2} are convex in $\boldsymbol{t}$ respectively. 

\begin{lemma} \label{Mconvex}
	The  constraints \eqref{M_telda_less_1_2} and \eqref{M_telda_less_1} are convex with respect to  $\boldsymbol{{t}}$. 
\end{lemma}
\begin{proof}
	The proof is provided in Appendix \ref{apdx:Mconvex}.
\end{proof}

\begin{lemma}
	The  constraints \eqref{eq:don_pos_cond} and \eqref{eq:don_pos_cond2} are convex with respect to   $\boldsymbol{{t}}$.  \label{don_convex}
\end{lemma}
\begin{proof}
	The proof is provided in Appendix \ref{apdx_don}.
\end{proof}

Algorithm \ref{alg:NOVA_Alg1} shows the used procedure to solve for $\boldsymbol{t}$. Let $\overline{U}\left(\boldsymbol{t};\boldsymbol{t^\nu}\right)$ be the
convex approximation at iterate $\boldsymbol{t^\nu}$ to the original non-convex problem $U\left(\boldsymbol{t}\right)$, where $U\left(\boldsymbol{t}\right)$ is given by (\ref{eq:joint_otp_prob}), assuming other parameters constant. Then, a valid choice of  $\overline{U}\left(\boldsymbol{t};\boldsymbol{t^\nu}\right)$ is the first order approximation of  $U\left(\boldsymbol{t}\right)$, i.e.,  
\begin{equation}
\overline{U}\left(\boldsymbol{t},\boldsymbol{t^\nu}\right)=\nabla_{\boldsymbol{t}}U\left(\boldsymbol{t^\nu}\right)^{T}\left(\boldsymbol{t}-\boldsymbol{t^\nu}\right)+\frac{\tau_{t}}{2}\left\Vert \boldsymbol{t}-\boldsymbol{t^\nu}\right\Vert ^{2}.\label{eq:U_t_u_bar}
\end{equation}
where $\tau_t$ is a regularization parameter. The detailed steps can be seen in Algorithm \ref{alg:NOVA_Alg1}. Since all the constraints \eqref{eq:t_i_alpha_j}, \eqref{M_telda_less_1},and  \eqref{eq:don_pos_cond} have been shown to be convex in $\boldsymbol{t}$, the optimization problem in Step 1 of  Algorithm \ref{alg:NOVA_Alg1} can be solved by the standard projected gradient descent algorithm. 



\subsection{Placement Optimization}
Given $\boldsymbol{\mathcal{\pi}}$ and $\boldsymbol{t}$, this subproblem finds a permutation of the placement of files on the different servers. Let the given $\boldsymbol{\mathcal{\pi}}$ be denoted as $\boldsymbol{\mathcal{\pi}}' = \{\pi'_{ij}\forall i,j\}$ and the placement corresponding to this access be $\boldsymbol{\mathcal{S}}'=\left(\mathcal{S}_{1}',\mathcal{S}_{2}',\ldots,\mathcal{S}_{r}'\right)$. We find a permutation of the servers $m$ for each file $i$, and call it $\zeta_i(j)$ is a permutation of the  servers from $j\in \{1,\cdots m\}$ to $\zeta_i(j) \in \{1,\cdots m\}$. Further, having the mapping of the servers for each file, the new access probabilities are $\pi_{ij}= \pi'_{i,\zeta_i(j)}$. Having these access probabilities, the new placement of the files will be $\mathcal{S}_{i} = \{\zeta_i(j) \forall j \in \mathcal{S}'_{i}\}$. We note that the constraints \eqref{eq:sum_ij},  \eqref{eq:pij}, and  \eqref{eq:S_i_and_ni} for the access from the modified placement of the servers will already be satisfied. The Placement Optimization subproblem is to find the optimal permutations $\zeta_i(j)$. The problem can be formally written as follows.



\textbf{Objective:} $\min$  \eqref{eq:joint_otp_prob}

\textbf{s.t.} \eqref{eq:rho_j}, \eqref{eq:Lambda_j}, \eqref{eq:don_pos_cond}, $\pi_{ij}= \pi'_{i,\zeta_i(j)}$,  ${\zeta_i}$ is a permutation on $\\\{1, \cdots, m\} \ \forall\   i\in \{1, \cdots, r\}$ 
 

\textbf{var.}  ${\zeta_i(j)} \quad  \forall j\in \{1, \cdots, m\}$ and $i\in \{1, \cdots, r\}$


We note that the optimization problem is to find $r$ permutations and is a discrete optimization problem. We first consider optimizing only  over one of the permutation $\zeta_i$. Let $\zeta_i$ be written as an indicator function $x_{u,v}^{(i)}$ which is $1$ if $v=\zeta_i(u) $ and zero otherwise. Then, the new $\pi_{ij} = \sum_u x_{j,u}^{(i)}\pi'_{iu}$ while for other files $k\ne i$, $\pi_{ij}$ remains the same. With the new values of $\pi_{ij}$, the only optimization variables are $x_{j,u}^{(i)}$. The constraints for  $x_{u,v}^{(i)}$ are $\sum_v x_{u,v}^{(i)} = \sum_u x_{u,v}^{(i)} =1$ and $x_{u,v}^{(i)}\in \{0,1\}$. We note that this is a non-linear bipartite matching problem \cite{Berstein200853}.  All the $r$ permutations taken together result in $rm^2$ discrete optimization variables that we wish to optimize. 

In general, we have the constraints $\pi_{ij} = \sum_u x_{j,u}^{(i)}\pi'_{iu}$ and $\sum_v x_{u,v}^{(i)} = \sum_u x_{u,v}^{(i)} =1$ for all $i\in \{1, \cdots, r\}$, $u, v \in \{1, \cdots, m\}$,  where binary $x_{u,v}^{(i)}$ for each $i, u, v$ are the decision variables.  In order to solve the non-linear problem with integer constraints, we use NOVA algorithm, where a term $\left(1+e^{\left(\alpha_{c}x\right)}\right)^{-1}-\left(1+e^{\left(\alpha_{c}\left(x-1\right)\right)}\right)^{-1}$ is added in the objective for each constraint (to make the problem smooth), where $\alpha_c$ is a large number and $C$ is large enough to force the solutions to be binary. NOVA algorithm guarantees convergence for any given value of $C$ and thus for large enough $C$, we will obtain the stationary point that has integer constraints.

\section{Proof of Lemma \ref{Mconvex}}\label{apdx:Mconvex}
The constraints \eqref{M_telda_less_1_2} and \eqref{M_telda_less_1} are separable for each $\widetilde{t}_i$ and $\overline{t}_i$ and thus it is enough to prove convexity of $C(t)=\alpha_{j}\left(e^{\left(\beta_{j}-\tau\right)t}-1\right)+t$. Thus, it is enough to prove that $C''(t)\ge 0$. 

The first derivative of $C(t)$ is given as

\begin{equation}
C'(t)=\alpha_{j}\left(\left(\beta_{j}-\tau\right)e^{\left(\beta_{j}-\tau\right)t}\right)+1
\end{equation}

Differentiating it again, we get the second derivative as follows. 
\begin{equation}
C''(t)  =\alpha_{j}\left(\beta_{j}-\tau\right)^{2}e^{\left(\beta_{j}-\tau\right)t}\label{eq:C_22}
\end{equation}

Since $\alpha_j>0$, $C''(t)$ given in  (\ref{eq:C_22}) is non-negative, which proves the Lemma.

\section{Proof of Lemma \ref{don_convex}} \label{apdx_don}
The constraints \eqref{eq:don_pos_cond} and \eqref{eq:don_pos_cond2}   are separable for each each $\widetilde{t}_i$ and $\overline{t}_i$, and thus it is enough to prove convexity of $E(t) = \sum_{f=1}^r\pi_{fj}\lambda_{f}\left(\frac{\alpha_{j}e^{\beta_{j}t}}{\alpha_{j}-t}\right)^{L_{f}}-\left(\Lambda_{j}+t\right)$ for $t<\alpha_j$. Thus, it is enough to prove that $E''(t)\ge 0$  for $t<\alpha_j$. We further note that it is enough to prove that $D''(t)\ge 0$, where $D(t) = \frac{e^{L_f\beta_{j}t}}{(\alpha_{j}-t)^{L_{f}}}$. Hence, the first derivative of $D(t)$ is given as

\begin{align}
D^{'}(t) & =\frac{L_{f}e^{L_{f}\beta_{j}t}\left[\beta_{j}+\left(\alpha_{j}-t\right)^{-1}\right]}{\left(\alpha_{j}-t\right)^{L_{f}}}>0
\end{align}

Note that $D'(t)>0$ since $\alpha_j>t$. 
Differentiating it again to get the second derivative, we get the second derivative as follows. 
\begin{eqnarray}
&&D^{''}(t)  =\frac{L_{f}\beta_{j}e^{L_{f}\beta_{j}t}}{\left(\alpha_{j}-t\right)^{L_{f}+2}}\times\nonumber\\
&& \left[\beta_{j}+\left(1+L_{f}\right)\left(\alpha_{j}-t\right)^{-1}\left(1+\frac{1}{\beta_{j}\left(\alpha_{j}-t\right)}\right)\right] \label{eq:E_22}
\end{eqnarray}




Since $\alpha_j>t$, $D''(t)$ given in  \eqref{eq:E_22} is non-negative, which proves the Lemma. 


	
\section{Additional Simulation Figures}\label{sec:par_enc}

In this section, in addition to the variations studied earlier, we will explore the effects of changing some other system parameters, i.e., the number of servers, the number of video files, the increase of video request arrival rates, and the code choice on the stall durations. 

\begin{figure}[t]
	\centering
	\includegraphics[trim=0.0in 0in 4.2in 0in, clip, width=0.48\textwidth]{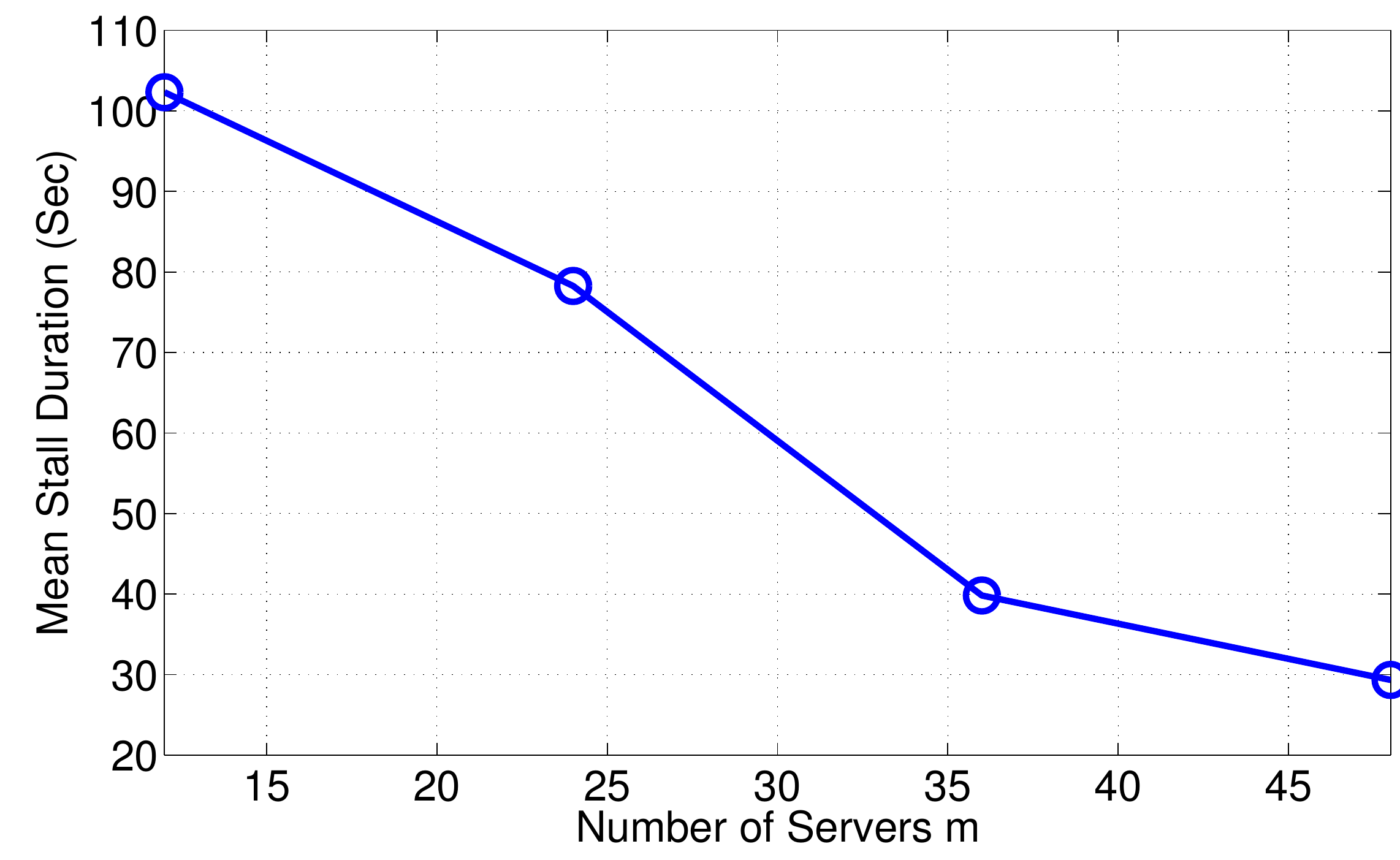}
	\caption{Mean stall duration for $2000$ files and different number of servers $m$ }
	\label{fig:meanStall_vs_Noserver}
\end{figure}

{\bf Effect of number of servers: } Figure \ref{fig:meanStall_vs_Noserver} depicts the mean stall duration for increasing number of servers ($12$, $24$, $36$, $48$). We note that the mean stall duration decreases with increase of servers. 

\begin{figure}[t]
	\centering
	\includegraphics[trim=0.01in 0in 4.2in 0in, clip,width=0.48\textwidth]{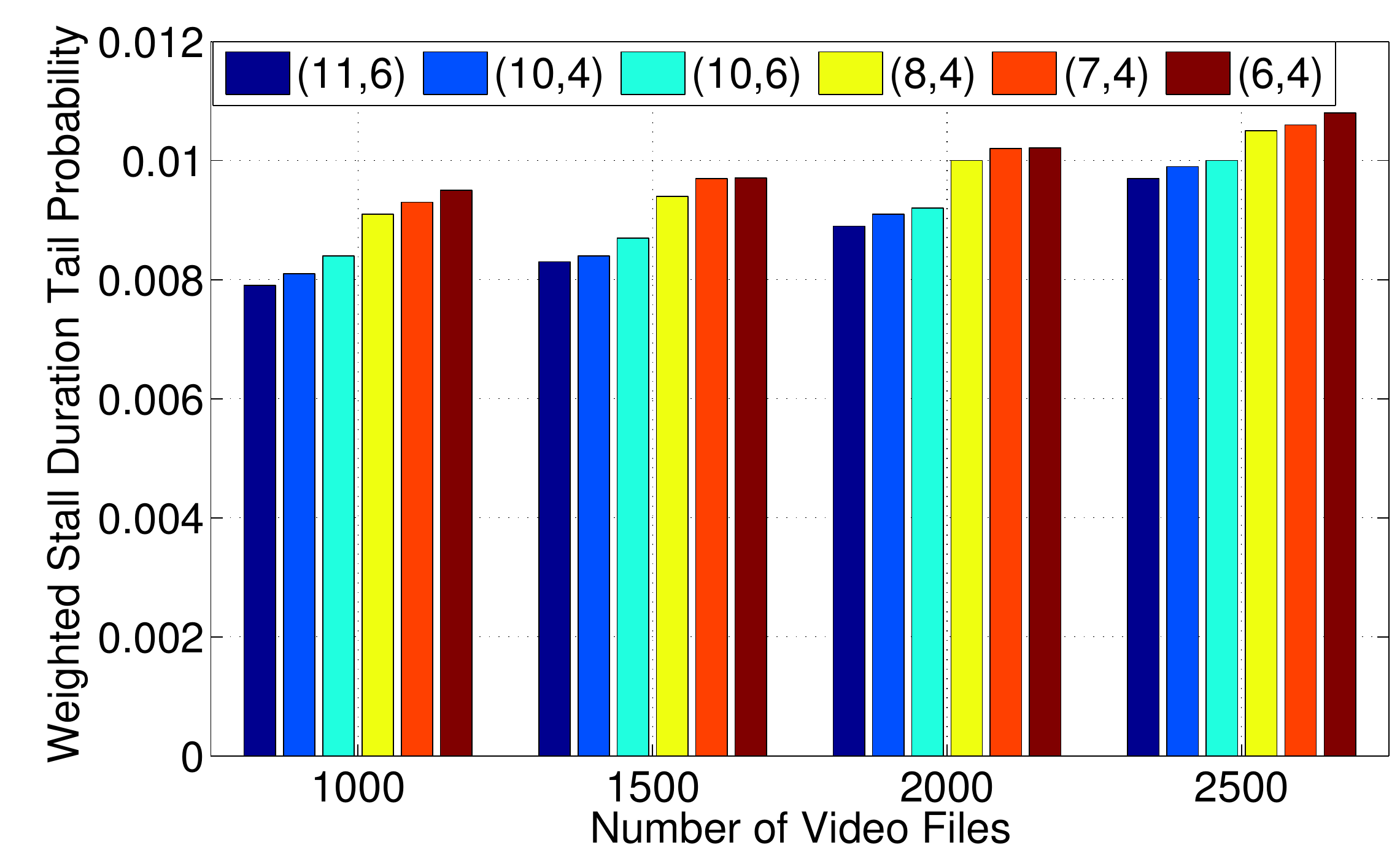}
	\caption{Weighted stall duration tail probability for different coding with different video lengths.}
	\label{fig:ersCodingEff}
\end{figure}

{\bf Effect of encoding parameters: } Figure \ref{fig:ersCodingEff} depicts the weighted stall duration tail probability for varying the number of files, and for different choices of code parameters. We first note that the weighted stall duration tail probability is higher for larger number of files. Further, we note that the code with larger $n$ for the same value of $k$ performs better. This is because larger value of $n$ gives more choice for the selection of servers. Thus, $(11,6)$ performs better than $(10,6)$ and $(8,4)$ performs better than $(7,4)$. Among $(10,6)$ and $(8,4)$, the additional redundancy is $4$. With the same number of parity symbols, it is better to have larger value of $k$ since smaller chunks are obtained from each server helping stall durations. Since the replication has $k=1$, this analysis thus shows that an erasure code with the same redundancy can help achieve better stall durations.

{\bf Performance with Repetition Coding: }
\begin{figure}[t]
	\centering
	\includegraphics[trim=0.01in 0in 4.0in 0in, clip,width=0.48\textwidth]{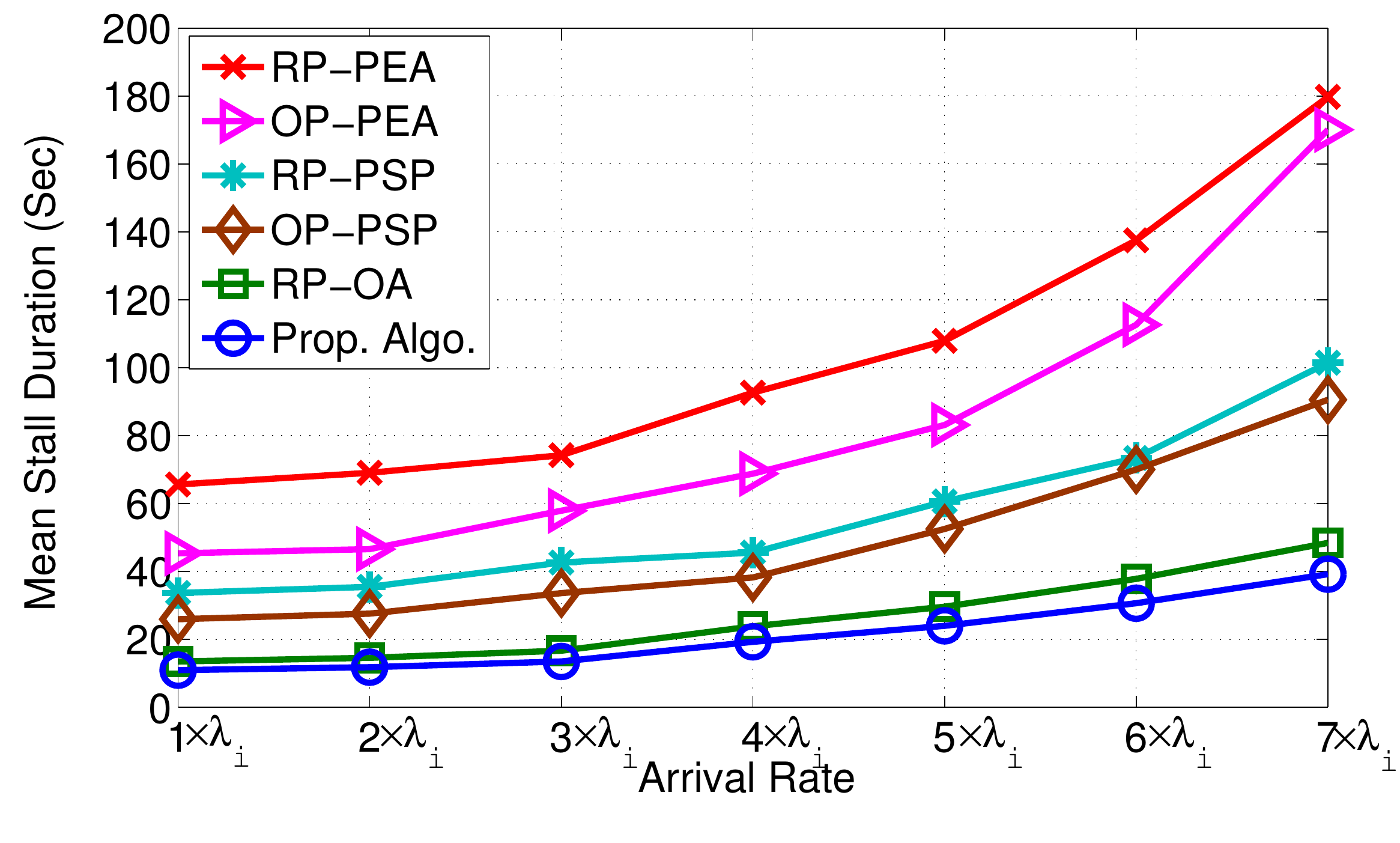}
	\caption{Mean Stall Duration for replication-based setup $(k=1)$. We set  $m=24$ servers, $r=2000$ video files, arrival rate is varied from $1\times \lambda_i$   to $7\times \lambda_i$, where $\lambda_i$ is the base arrival rate. The video file sizes are Pareto-based distributed, i.e., can be anywhere between 1-120 minutes.}
	\label{fig:repCoding}
\end{figure} 
Figure \ref{fig:repCoding}
shows the effect of different video arrival rates on the mean stall duration for different-size video length when each file uses $(3,1)$ erasure-code (which is triple-replication).  We compare our proposed algorithm with the five baseline  policies and see that the proposed algorithm outperforms all baseline strategies for the QoE metric of mean stall duration. Thus, both access and placement of files are  important for reducing the mean stall duration. We see that the mean stall duration of all approaches increases with arrival rates. However, since the mean stall duration is more significant at high arrival rates, we see the significant improvement in the mean stall duration  of our approach as compared to the considered baselines.

\begin{figure}[t]
	\centering
	\includegraphics[trim=0.1in 0in 4.3in 0in, clip, width=0.42\textwidth]{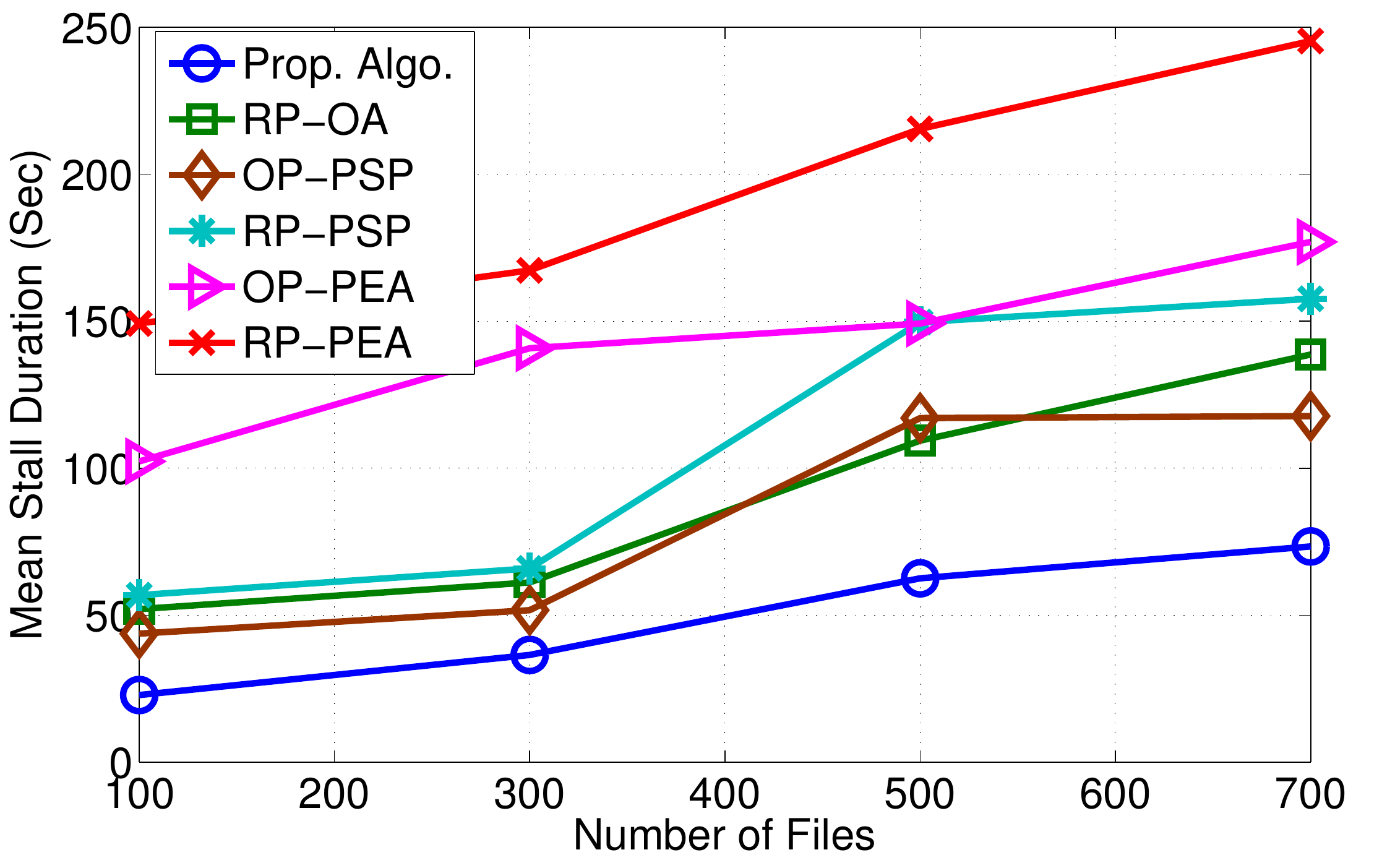}
	\captionof{figure}{Mean stall duration for different number of video files with different video lengths.}
	\label{fig:meanStallNumFiles} 
\end{figure}%

\begin{figure}[t]
	\centering
			\includegraphics[trim=0in 0.1in 4.0in 0in, clip, width=0.42\textwidth]{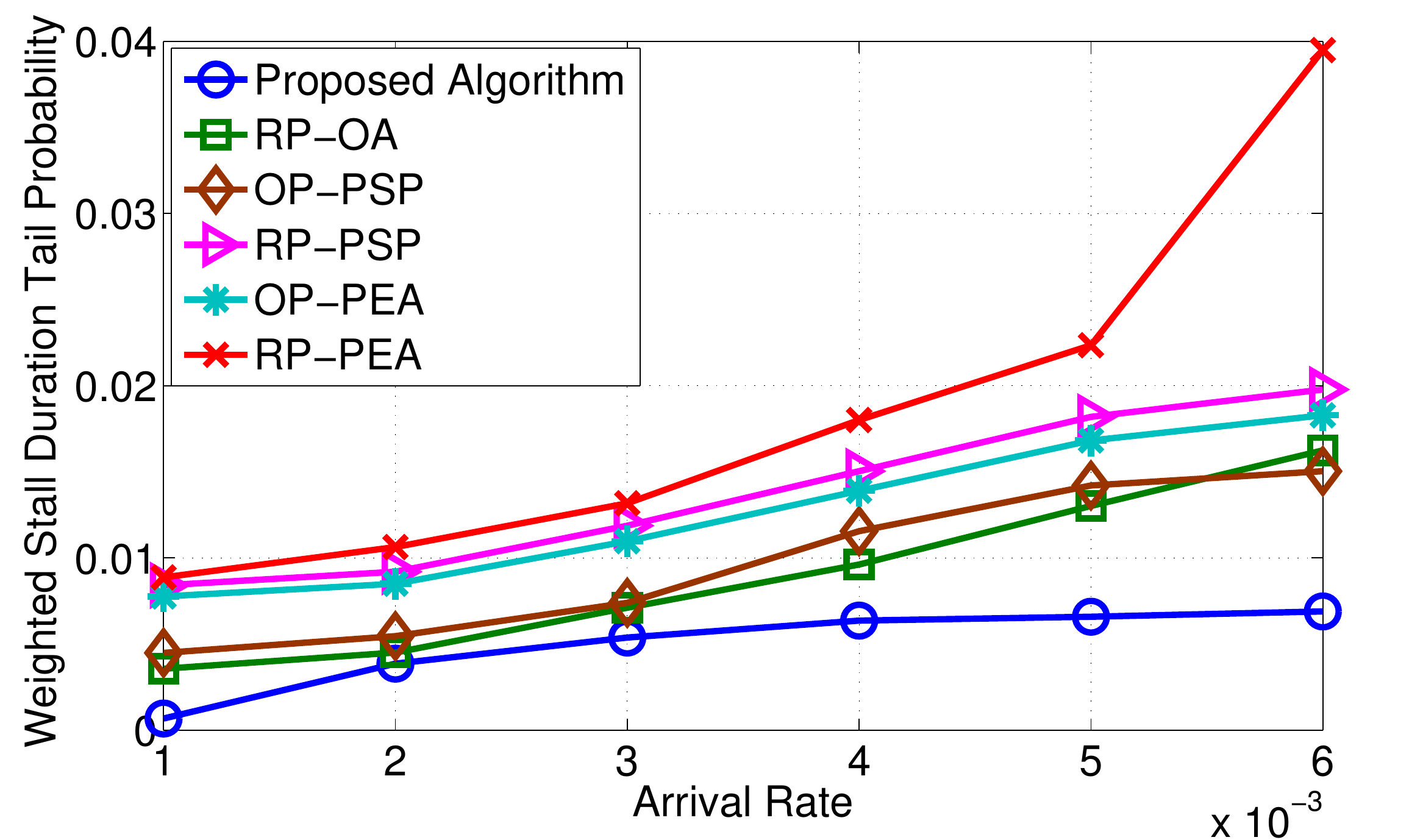}
			\captionof{figure}{Stall duration tail probability for different arrival rates for video files ($x=150$ s).}
			\label{fig:tailProbArrRate}
\end{figure}

{\bf Effect of Arrival Rates} Figure \ref{fig:tailProbArrRate} demonstrates the effect of increasing  workload, obtained by varying the arrival rates of the video files from $0.25\lambda$ to $2\lambda$, where $\lambda$ is the base arrival rate, on the stall duration tail probability for video lengths generated based on Pareto distribution defined above. We notice a significant improvement of the QoE metric with the proposed strategy as compared to the baselines. At the arrival rate of $2\lambda$, the proposed strategy reduces the stall duration tail probability by about 100\% as compared to the random placement and projected equal access policy. 

\begin{figure}[t]
	\centering
	\includegraphics[trim=0.1in 0in 4.4in 0in, clip, width=0.4\textwidth]{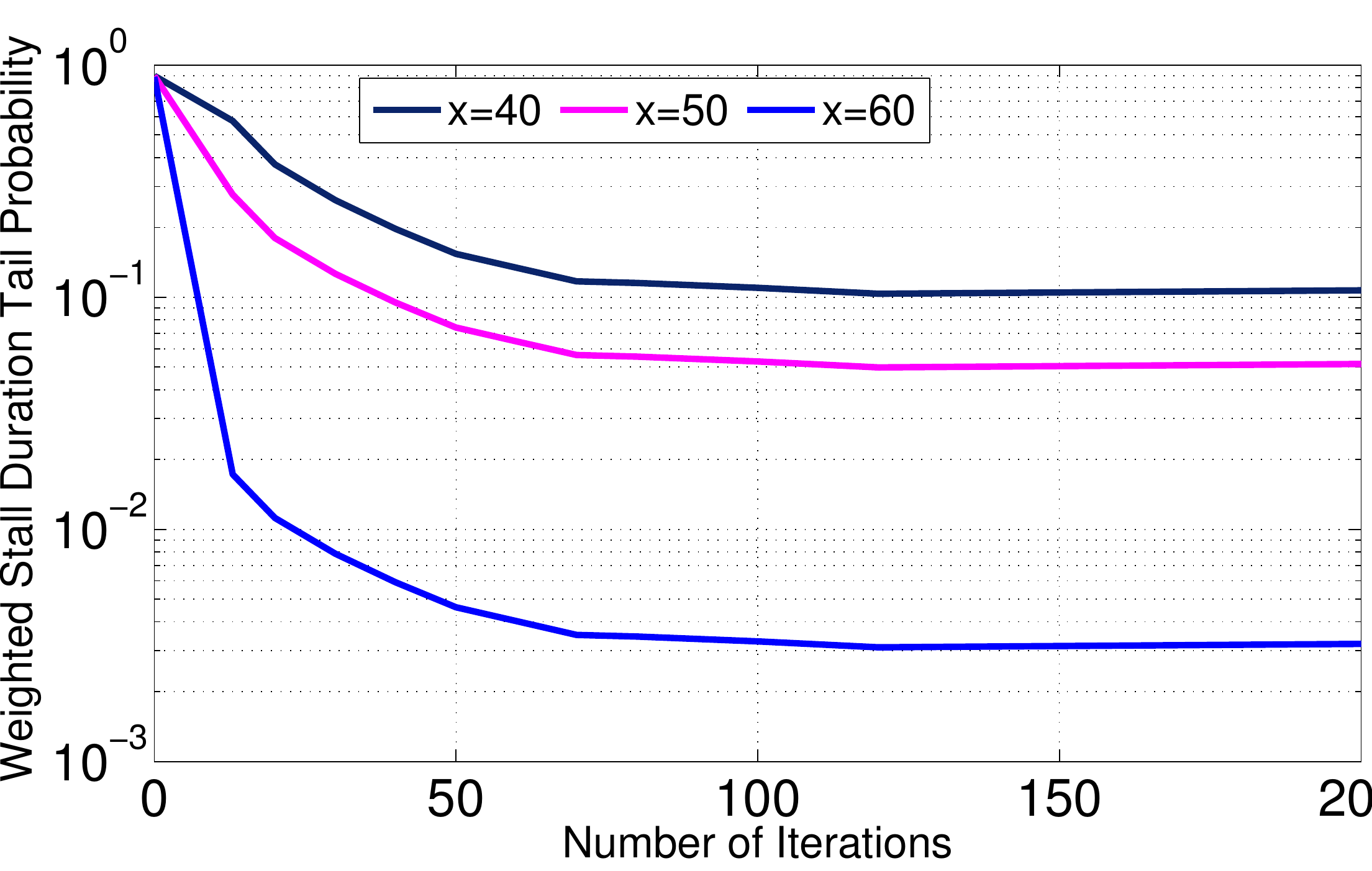}
	\captionof{figure}{Stall Duration Tail Probability for different number of iterations. }
	\label{fig:congStallDiffXvalues}
\end{figure}

{\bf Convergence of Stall Duration Tail Probability}
Figure \ref{fig:congStallDiffXvalues} demonstrates the convergence  of our proposed algorithm for different values of $x$. Considering $r=1000$ files of length 300s each with $m=12$ storage nodes, the  stall duration tail probability converges to the optimal value within less than $200$ iterations.

{\bf Effect of the Number of Video Files}
Figure \ref{fig:meanStallNumFiles} demonstrates the impact of varying the number of video files from $100$ files to $700$ files  on the mean stall duration, where the video lengths are generated according to Pareto distribution with the same parameter defined earlier (scale of 300, and shape of 2).  We note that the proposed optimization strategy effectively reduces the mean stall duration and outperforms the considered baseline strategies. Thus, joint optimization over all three variables $\boldsymbol{\mathcal{S}}$,  $\boldsymbol{\pi}$, and  $\boldsymbol{t}$ helps reduce the mean stall duration significantly.
\section{Extension to more streams between the server and the edge router} \label{apdx_entend}

\begin{figure}
\includegraphics[trim=0.0in 4.5in 0.1in .1in, clip, width=.45\textwidth]{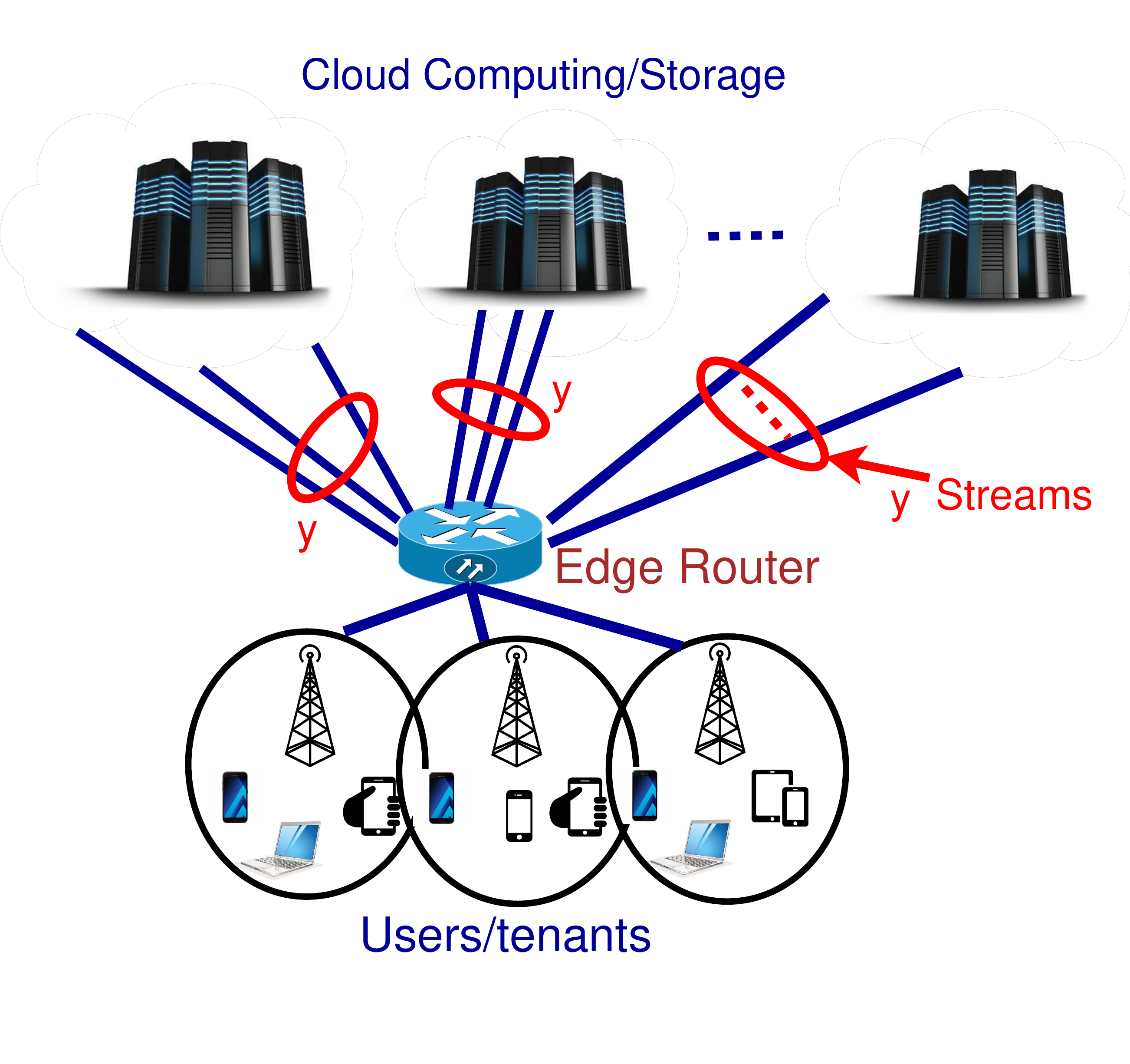}
\vspace{-.3in}
\caption{An Illustration of a distributed storage system where a server has $y$ parallel streams to the edge router.\label{fig:PS_y}}
\vspace{-.2in}
\end{figure}

 In this section, we investigate  extending the proposed approach to the case when there are $y$ parallel streams from each server to the edge router. Multiple streams can help obtain parallel video files thus helping one file not wait behind the other. We label the $y$ streams from server $j$ as $\nu_j \in \{1, \cdots, y\}$ (graphically depicted in Figure  \ref{fig:PS_y}).  The  analysis in this paper considers only one stream between the server and the edge router. We now show how the analysis can be adapted when there are multiple streams. We first note that the scheduling need to decide not only the server $j$ but also the parallel stream $\nu_j$. We assume that the parallel stream $\nu_j$ is chosen equally likely. Further, the multiple streams are obtained through equal bandwidth splits, and thus the service time parameters would be different for streams as compared to the server. For instance, the service rate would be a factor of $y$ of the service rate from the server due to the bandwidth split. Thus, the probability of choosing server $j$ and stream $\nu_j$ is 
	\begin{equation}
	q_{i,j,\nu_{j}}=\pi_{i,j}/y,\label{eq:pi_i_j_nu}
	\end{equation}
	where $\pi_{i,j}$ is the probability of choosing server $j$. Using this, we note that the analysis of download time from a server can be modified to download time from a stream of a server and the steps can be directly extended. The ordered statistics can use the above probabilistic scheduling to choose a stream of a server and thus the entire analysis can be easily extended. 
	
\begin{figure}[t]
		\centering
			\includegraphics[trim=0in 0in 4.3in 0in, clip, width=0.48\textwidth]{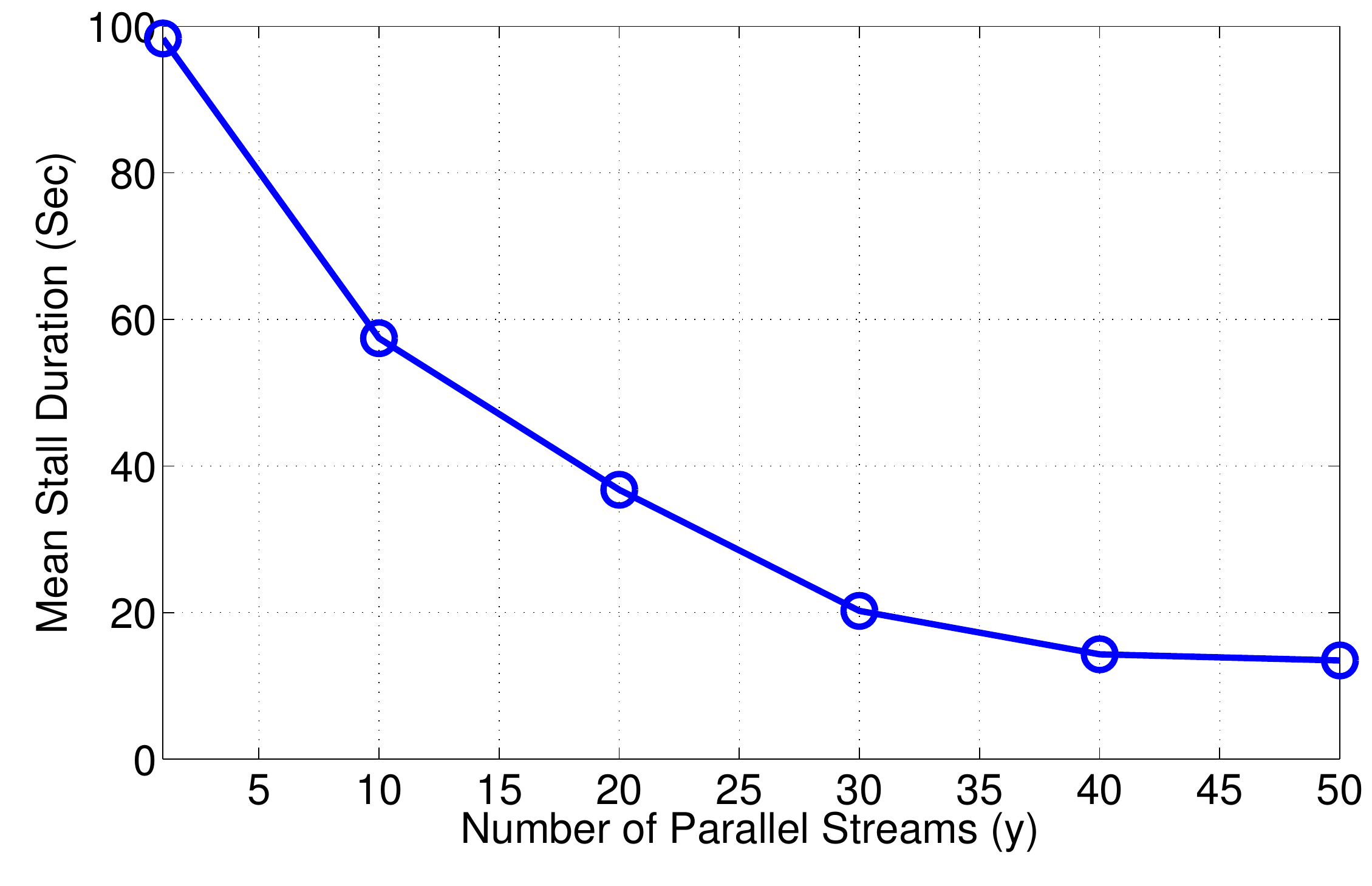}
			\caption{Mean stall duration for different number of parallel streams.}
			\label{fig:meanStall_vs_y_j}
		\end{figure}

	Since the optimization also has the same parameters, we show  an improvement of the mean stall duration with the number of parallel streams in Fig. \ref{fig:meanStall_vs_y_j}. The choice of the number of streams $y$ can be determined by the practical limitations ({\it e.g.}, number of ports possible at the server). A more detailed analysis of the parallel streams, exploiting the flexibilities of  splitting of bandwidths among the different streams, choosing one of the multiple parallel streams for each video are being considered by the PI in \cite{Abubakr_SPCOM, Abubakr_Stall_Qual}.

\section{Future Directions}\label{apd:future}

A server does not need to serve different video requests one after the other. It may be better to serve video segments out of order from a queue thus helping stall durations since the later video requests do not have to wait for finishing chunks of earlier requests which have later deadlines. Exploiting this flexibility is an open problem. 

Most edge routers would have a cache capacity, where certain segments of video files can be stored to improve the QoEs. Analyzing QoEs and finding efficient caching mechanisms for video streaming over cloud  is an open problem (See Appendix \ref{apdx:cache} for more details). 
%
Further, implementing the ideas in this paper over a real cloud computing environment is left as a future work. This paper also does not consider the last hop, and incorporating that is left as a future work (See Appendix \ref{apdx:e2e} for more details). 

We note that the current video streaming algorithms use adaptive bit-rate (ABR) strategies to change the video qualities of segments within a video \cite{huang2015buffer,AnisTON}. One of the strategies look at the buffer usage at the client to determine the quality of the next segment \cite{huang2015buffer}. Incorporating efficient ABR streaming algorithms is an interesting future work. The main challenge in this extension is to incorporate the client behavior which makes the  arrival process non-memoryless thus making the analysis complex.  Finally, considering the decoding time by combining data in the calculations is left as a future work.
\section{Impact Of Caching} \label{apdx:cache}

So far, our  analysis did not account for caching. In this Appendix, we present how our model can be extended to accommodate for the impact of caching. Caching content  at the network edge, closer to the customers, can  further help reducing the stall duration and thus improve the QoE. However, caching the video content has to address a number of crucial
challenges that differ from caching of web objects, see for
instance \cite{cache16} and the references therein for detailed treatment of this aspect.

There are two methods for caching the video files. The first involves caching the complete video file (all $L_i$ video chunks of file $i$) at edge routers. The second method involves caching partial chunks, i.e., $L_{j,i}$, where $L_{j,i}\leq L_{i}$, for video file $i$. Most of the current caching schemes cache entire files (for example, hot files).  Our analysis can accommodate both of these methods. In the first method, the video file is entirely cached, and is thus not requested from the servers. This is equivalent to  changing the arrival rate of these files to zero, i.e., $\lambda_i=0$. In the second method, only the later $(L_{i} - L_{j,i})$ are needed from the servers. This can be easily incorporated by requesting the video of length $(L_{i} - L_{j,i})$, while the first chunk can wait for an additional $\tau L_{j,i}$ time which can be accounted by adding $\tau L_{j,i}$ in the startup delay for this file.


Mathematically, we can show that for $g \in{L_{j,i}+1,\cdots,L_i}$, where $L_{j,i}<g\le L_{i}$, the random download time of the remaining $(L_i-L_{j,i})$ segments from server $j$ is given as 
 \begin{equation}
D_{i,j}^{(g)}=W_{j}+\sum_{v=L_{j,i}+1}^{g}Y_{j}^{(v)}.
\end{equation}
Since video file $i$ consists of $(L_i-L_{j,i})$ segments stored at
server $j$, the total service time for video file $i$, denoted by $ST_{i,j}$, is given as 
\begin{equation}
ST_{i,j}=\sum_{v=L_{j,i}+1}^{L_{i}}Y_{i,j}^{(v)}
\end{equation}
Hence, the service time of the video files at server $j$ is given by  
\begin{equation}
R_{j}=\begin{cases}
ST_{i,j} \ \   \text { with prob. } \frac{\lambda_{i}\pi_{i,j}}{\Lambda_{j}}& \forall i\end{cases}
\end{equation}
where $\Lambda_{j}$ is the total arrival rate at server $j$. Also, we can show that the MGF of the service time for all video files from server $j$ is given by 
\begin{equation}
B_{j}(t)={\mathbb E}[e^{tR_{j}}] =\sum_{i=1}^{r}\frac{\lambda_{i}\pi_{i,j}}{\Lambda_{j}}\left(\frac{\alpha_{j}e^{\eta_{j}t}}{\alpha_{j}-t}\right)^{(L_i-L_{j,i})}\label{Be}
\end{equation}
Further, the current load intensity  at server $j$, $\rho_j$, is as follows 
\begin{equation}
\rho_{j}  = \Lambda_{j} B_{j}'(0)=  \sum_{i=1}^{r}\lambda_{i}\pi_{i,j}(L_i-L_{j,i})\left( \eta_{j}+\frac{1}{\alpha_j}\right) \label{rhoe}
\end{equation}

Similar to the previous analysis, since the arrival is Poisson and the service time is shifted-exponentially distributed, the MGF of the waiting time at queue server $j$ can be calculated usingthe  Pollaczek-Khinchine formula, i.e.,
\begin{align}
\mathbb{E}\left[e^{tW_j}\right] & =\frac{(1-\rho_{j})tB_{j}(t)}{t-\Lambda_{j}(B_{j}(t)-1)}
\end{align}
From the MGF of $W_{j}$ and the service time, the MGF of the download time of segment $g$ from server $j$ for file $i$ is then
\begin{equation}
\mathbb{E}\left[e^{tD_{i,j}^{(g)}}\right]=\frac{(1-\rho_{j})tB_{j}(t)}{t-\Lambda_{j}(B_{j}(t)-1)}\left(\frac{\alpha_{j}e^{\eta_{j}t}}{\alpha_{j}-t}\right)^{g}
\label{D_nu_j}.
\end{equation}
Having characterized the download time for chunk $g$, we can then determine the stall durations and evaluate the QoE metrics as in Equations \eqref{eq:T_s_main_stall} and \eqref{eq:T_stall}. The details are omitted here as they can easily follow. 
\section{End-to-End Analysis} \label{apdx:e2e}

In this Appendix, we show how our analysis can be extended to consider the last hop from the edge-router to the user. If the last hop is considered, the download time of the chunk $q$ for video file $i$, if requested from server $j$ can be written as follows

\[
D_{i,j}^{(q)}=W_{j}+\sum_{v=1}^{q}Y_{j}^{(v)}+\frac{\tau\mathbb{R}_{i}}{\mathbb{C}_{i}},
\]
where $W_j$ is the waiting time in the queue of server $j$, $Y_{j}^{(v)}$ is the service time for the chunk $v$, $\tau$ is the chunk size in seconds, $\mathbb{R}_{i}$ is the bit-rate  for user $i$, and $\mathbb{C}_{i}$ is the average bandwidth when downloading  chunk $q$. Thus, as long as $\frac{\tau\mathbb{R}_{i}}{\mathbb{C}_{i}}$ can be bounded, this is the additional stall duration (or additional startup delay). In most wired setups, the capacity for the last hop may not be a bottleneck, and thus this term is negligible and not varying significantly with $q$. Even for wireless network in homes, the average bandwidth numbers are much higher than the video rate, and thus this additional term may not be a bottleneck. Thus, the analysis can be easily extended to the last hop. Since the last hop is dependent on the user and the cloud provider wishes to optimize the system such that it does the best delivery in the part controlled by the provider, we did not explicitly consider the last hop. However, as long as the last hop capacity is higher than the data rate of the video, the last hop does not affect the analysis except a small additional delay.

\end{document}